\documentclass[lettersize,journal]{IEEEtran}
\usepackage{amsmath,amsfonts}
\usepackage{array}
\usepackage[caption=false,font=normalsize,labelfont=sf,textfont=sf]{subfig}
\usepackage{textcomp}
\usepackage{stfloats}
\usepackage{url}
\usepackage{verbatim}
\usepackage{graphicx}
\usepackage{cite}

\usepackage{booktabs}       
\usepackage{nicefrac}       
\usepackage{microtype}      
\usepackage{xcolor}         
\usepackage{xspace}
\usepackage{wrapfig}
\usepackage{amssymb}
\usepackage{bm}
\usepackage{tabularx}
\usepackage{paralist}
\usepackage{mathtools}
\usepackage[linesnumbered,ruled,vlined]{algorithm2e}
\usepackage{pifont}
\usepackage{multirow}

\makeatletter
\DeclareRobustCommand\onedot{\futurelet\@let@token\@onedot}
\def\@onedot{\ifx\@let@token.\else.\null\fi\xspace}
\def\eg{\emph{e.g}\onedot} 
\def\ie{\emph{i.e}\onedot} 
 
\def\etc{\emph{etc}\onedot} \def\vs{\emph{vs}\onedot}
 
\def\etal{\emph{et al}\onedot}
\makeatother

\newcommand{\Tref}[1]{Table~\ref{#1}}
\newcommand{\Eref}[1]{Equation~(\ref{#1})}
\newcommand{\Fref}[1]{Figure~\ref{#1}}

\newcommand{\fref}[1]{Fig.~\ref{#1}}

\newcommand{\Frst}[1]{\textcolor{red}{\textbf{#1}}}
\newcommand{\Scnd}[1]{\textcolor{blue}{\textbf{#1}}}

\begin{document}

\title{Architectural Unification for Polarimetric Imaging Across Multiple Degradations}

\author{
Chu Zhou, Yufei Han, Junda Liao, Linrui Dai, Wangze Xu, Art Subpa-Asa, Heng Guo, Boxin Shi,~\IEEEmembership{Senior Member,~IEEE} and Imari Sato,~\IEEEmembership{Member,~IEEE}
    \thanks{
    Chu Zhou, Junda Liao, Linrui Dai, Wangze Xu, Art Subpa-Asa, and Imari Sato are with the Digital Content and Media Sciences Research Division, National Institute of Informatics, Tokyo 101-8430, Japan.
    }
    \thanks{
    Yufei Han and Heng Guo are with the Pattern Recognition and Intelligent System Laboratory, School of Artificial Intelligence, Beijing University of Posts and Telecommunications, Beijing 100876, China
    }
    \thanks{
    Boxin Shi is with the State Key Laboratory of Multimedia Information Processing, School of Computer Science, Peking University, Beijing 100080, China, and also with the National Engineering Research Center of Visual Technology, School of Computer Science, Peking University, Beijing 100080, China.
    }
}

\markboth{IEEE TRANSACTIONS ON PATTERN ANALYSIS AND MACHINE INTELLIGENCE}%
{Shell \MakeLowercase{\textit{et al.}}: A Sample Article Using IEEEtran.cls for IEEE Journals}

\IEEEpubid{0000--0000/00\$00.00~\copyright~2021 IEEE}

\maketitle

\begin{abstract}
Polarimetric imaging aims to recover polarimetric parameters, including Total Intensity (TI), Degree of Polarization (DoP), and Angle of Polarization (AoP), from captured polarized measurements. In real-world scenarios, these measurements are frequently affected by diverse degradations such as low-light noise, motion blur, and mosaicing artifacts. Due to the nonlinear dependency of DoP and AoP on the measured intensities, accurately retrieving physically consistent polarimetric parameters from degraded observations remains highly challenging. Existing approaches typically adopt task-specific network architectures tailored to individual degradation types, limiting their adaptability across different restoration scenarios. Moreover, many methods rely on multi-stage processing pipelines that suffer from error accumulation, or operate solely in a single domain (either image or Stokes domain), failing to fully exploit the intrinsic physical relationships between them. In this work, we propose a unified architectural framework for polarimetric imaging that is structurally shared across multiple degradation scenarios. Rather than redesigning network structures for each task, our framework maintains a consistent architectural design while being trained separately for different degradations. The model performs single-stage joint image-Stokes processing, avoiding error accumulation and explicitly preserving physical consistency. Extensive experiments show that this unified architectural design, when trained for specific degradation types, consistently achieves state-of-the-art performance across low-light denoising, motion deblurring, and demosaicing tasks, establishing a versatile and physically grounded solution for degraded polarimetric imaging.
\end{abstract}


\section{Introduction}
\IEEEPARstart{P}{olarization} is an intrinsic property of light, alongside amplitude and phase. It conveys rich physical information through polarimetric parameters including Total Intensity (TI), Degree of Polarization (DoP), and Angle of Polarization (AoP), which have demonstrated significant potential in diverse downstream applications, including shape estimation \cite{deschaintre2021deep, lyu2023shape}, reflection removal \cite{lei2020polarized, lyu2022physics, yao2025polarfree}, dehazing \cite{schechner2001instant, zhou2021learning}, transparent object segmentation \cite{kalra2020deep, mei2022glass}, \etc. Since the reliability of these applications critically depends on the fidelity of the recovered physical quantities, obtaining high-quality polarimetric parameters under real-world conditions is of substantial importance.

Polarimetric imaging seeks to recover the polarimetric parameters, typically via intermediate Stokes representations, from captured polarized measurements. However, in practical scenarios, polarized observations are often degraded by adverse conditions such as low-light noise, motion blur, or mosaicing artifacts. Due to the nonlinear dependency of DoP and AoP on measured intensities, even moderate degradations can lead to significant distortions in the estimated physical parameters. Designing robust restoration methods for degraded polarimetric measurements therefore remains a fundamental challenge.

Existing polarization-aware restoration methods are typically developed for a single degradation type, with architectures highly specialized to that scenario \cite{hu2020iplnet, xu2022colorpolarnet, zhou2025learning, zhou2025pidsr}. When retrained or adapted to other degradation types, these specialized designs often exhibit limited effectiveness, suggesting a lack of architectural versatility. While general-purpose RGB restoration models \cite{zamir2022restormer, wang2022uformer} can be trained for different degradations using a shared backbone, they are not polarization-aware and fail to explicitly model the physical relationships. From a practical standpoint, a unified architecture is highly desirable: real-world polarimetric systems frequently encounter diverse and unpredictable degradations, making the deployment of multiple highly engineered, task-specific models computationally prohibitive and inflexible. These observations raise an important question: \emph{Can we design a polarization-aware architecture that remains structurally consistent across different degradation scenarios, while maintaining strong restoration performance without redesigning task-specific pipelines for each case?} 

\IEEEpubidadjcol

\begin{figure*}[t]
    \centering
    \includegraphics[width=1.0\linewidth]{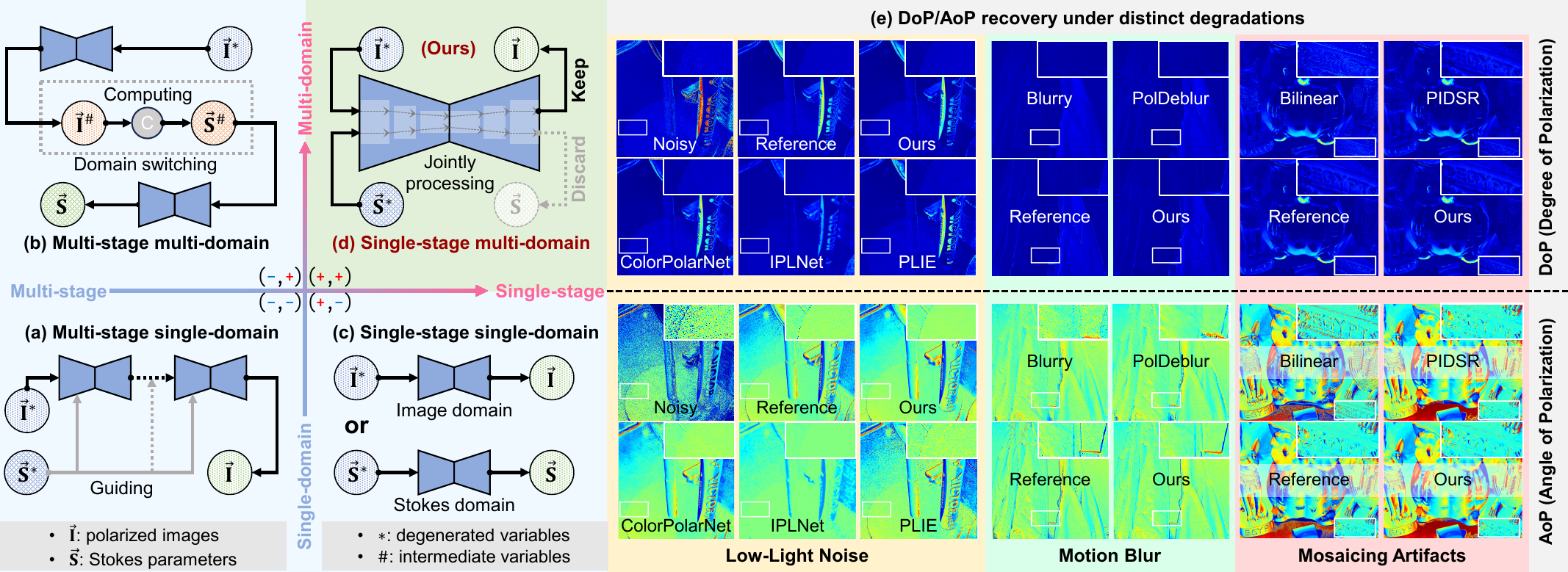}
    \caption{Comparison of inference pipelines (\Scnd{multi-stage} \vs \Frst{single-stage}) and representation domains (\Scnd{single-domain} \vs \Frst{multi-domain}) in polarimetric imaging. Existing approaches occupy three quadrants: (a) (\Scnd{-}, \Scnd{-}) multi-stage single-domain methods (\eg, PolDeblur \cite{zhou2025learning} and PIDSR \cite{zhou2025pidsr}); (b) (\Scnd{-}, \Frst{+}) multi-stage multi-domain methods (\eg, ColorPolarNet \cite{xu2022colorpolarnet}); (c) (\Frst{+}, \Scnd{-}) single-stage single-domain methods (\eg, IPLNet \cite{hu2020iplnet} and PLIE \cite{zhou2023polarization}). (d) (\Frst{+}, \Frst{+}) Our framework fills the previously unexplored first quadrant, providing a single-stage multi-domain architecture. (e) Experimental results demonstrating that our unified architecture, when optimized for distinct tasks, consistently achieves superior DoP/AoP recovery across diverse degradation scenarios, including low-light denoising, motion deblurring, and demosaicing; please zoom in for better details.}
    \label{fig: Teaser}
\end{figure*}

To answer this question, we systematically examine existing polarimetric restoration methods along two orthogonal design dimensions: (i) the \textit{inference pipeline} (\Scnd{multi-stage} \vs \Frst{single-stage}), and (ii) the \textit{representation domain} (\ie, the raw intensity image domain and the Stokes parameter domain; \Scnd{single-domain} \vs \Frst{multi-domain}). This yields a four-quadrant design space. As illustrated in \Fref{fig: Teaser}, prior methods occupy three of these quadrants:
\begin{itemize}
    \item[(\Scnd{-}, \Scnd{-})] \textit{Multi-stage single-domain} methods (\eg, PolDeblur \cite{zhou2025learning}, PIDSR \cite{zhou2025pidsr}), which decompose restoration into multiple sequential steps within the image domain (\Fref{fig: Teaser} (a)).
    \item[(\Scnd{-}, \Frst{+})] \textit{Multi-stage multi-domain} methods (\eg, ColorPolarNet \cite{xu2022colorpolarnet}), which involve both image and Stokes representations but still rely on staged processing (\Fref{fig: Teaser} (b)).
    \item[(\Frst{+}, \Scnd{-})] \textit{Single-stage single-domain} methods (\eg, IPLNet \cite{hu2020iplnet}, PLIE \cite{zhou2023polarization}), which adopt end-to-end learning yet operate in only one representation domain (\Fref{fig: Teaser} (c)).
\end{itemize}
Multi-stage pipelines often suffer from increased complexity and potential error accumulation across stages, whereas single-domain designs may fail to fully capture the intrinsic physical coupling between the image and Stokes domains. Notably, the \textit{single-stage multi-domain} quadrant (\Frst{+}, \Frst{+}) remains largely unexplored. Filling this missing quadrant offers the potential to simultaneously enable end-to-end optimization and explicit cross-domain physical modeling.

In this work, we propose a unified architectural framework for polarimetric imaging that achieves single-stage multi-domain processing, occupying this previously unexplored quadrant, as shown in \Fref{fig: Teaser} (d). Distinct from existing polarization-aware methods whose network structures are tightly coupled to specific degradation types, our framework achieves architectural unification across degradations. The proposed architecture remains structurally fixed, serving as a versatile backbone that can be instantiated and trained for different degradation scenarios without requiring task-specific structural redesign. At the core of our framework are Cross-Domain Collaborative Interaction (CDCI) units, which facilitate joint image-Stokes processing within a single-stage pipeline. By enabling collaborative feature aggregation and cross-domain modulation, the architecture inherently preserves physical consistency while avoiding the error accumulation associated with multi-stage designs. To summarize, our contributions are three-fold:
\begin{itemize}
    \item[$\bullet$] We introduce a unified architectural framework for polarimetric imaging that achieves structural consistency across multiple degradation scenarios. This fundamentally addresses the limitations of existing polarization-aware approaches that rely on heavily engineered, degradation-specific network designs.
    \item[$\bullet$] We propose a single-stage, multi-domain processing paradigm that fills a critical gap in the polarimetric restoration design space. Driven by our CDCI units, this paradigm effectively exploits the synergistic relationship between the image and Stokes domains while enforcing explicit physical consistency.
    \item[$\bullet$] We demonstrate that the proposed unified architecture achieves state-of-the-art performance across diverse polarimetric imaging challenges. Extensive experiments validate its effectiveness on low-light noise, motion blur, and mosaicing artifacts when the same architectural backbone is optimized for each specific task.
\end{itemize}

\section{Related Work}
\subsection{Polarization-based Vision}
Polarization-based vision seeks to overcome inherent limitations of conventional RGB imaging by exploiting additional physical cues encoded in polarimetric parameters. These cues provide complementary information about scene geometry, material properties, and light transport mechanisms, enabling enhanced perception capabilities across various fields.

In robotics and autonomous systems, polarization has proven particularly valuable for handling challenging materials such as transparent or specular objects, whose appearance often violates Lambertian assumptions. The distinctive polarimetric responses of such objects have been leveraged for transparent object segmentation \cite{kalra2020deep, mei2022glass}, road surface detection \cite{li2020full}, and broader scene understanding tasks \cite{liang2022multimodal}. In computer vision, polarization offers rich constraints for 3D reasoning. Because the polarization state of reflected light is closely related to surface normals, micro-geometry, and material characteristics, it has been incorporated into shape estimation \cite{deschaintre2021deep, lyu2023shape}, inverse rendering \cite{dave2022pandora, han2024nersp, han2025polgs}, and depth sensing frameworks \cite{kadambi2017depth, tian2023dps}. Beyond geometric reconstruction, polarization has also been used to analyze light transport phenomena. Changes in polarization induced by reflection and scattering provide cues for reflection removal \cite{lyu2022physics, lei2020polarized, yao2025polarfree}, image dehazing \cite{schechner2001instant, zhou2021learning}, and underwater visibility enhancement \cite{schechner2006recovery, shen2024polarization}. Moreover, polarization properties exhibit invariances that can be exploited for photometric tasks. For instance, the DoP of achromatic surfaces remains achromatic under varying illumination conditions, which can assist color constancy estimation \cite{ono2022degree}. The spatially varying attenuation introduced by polarizers further enables high dynamic range imaging \cite{ting2021deep, zhou2023polarizationHDR}. Differences in polarization characteristics between direct sunlight and skylight have also been explored for shadow removal \cite{zhou2025polarization}. 

Collectively, these advances highlight the broad applicability of polarization cues across perception tasks. Importantly, many of these downstream algorithms are highly sensitive to inaccuracies in the estimated polarimetric parameters. This sensitivity underscores the necessity of recovering physically consistent and high-fidelity polarimetric information, particularly under degraded imaging conditions.

\subsection{Handling Degradations in Polarimetric Imaging}
Robust polarimetric imaging under degraded conditions has attracted increasing attention. Several works have explored computational photography frameworks \cite{zhou2024quality, zhou2026towards} and neural representations \cite{peters2023pcon, zhou2025polarimetric} to improve reconstruction quality. These approaches typically leverage modified capture settings, complementary acquisition strategies, or per-scene optimization. While effective under controlled settings, their reliance on specific acquisition protocols or scene-dependent training limits practical deployment in general restoration scenarios.

For convenience and flexibility, most existing studies adopt post-processing paradigms tailored to particular degradation types. In the case of low-light noise, IPLNet \cite{hu2020iplnet} employed residual dense blocks to jointly process multiple polarized low-light inputs. PLIE \cite{zhou2023polarization} introduced a dual-branch architecture operating in the Stokes domain to mitigate enhancement artifacts. ColorPolarNet \cite{xu2022colorpolarnet} and SGLE-Net \cite{dong2025sgle} further proposed two-stage pipelines that sequentially refine representations in both the image and Stokes domains. Lu \etal \cite{lu2024polarization} designed a four-stage enhancement network specifically for nighttime polarized imaging. For motion blur, PolDeblur \cite{zhou2025learning} adopted a divide-and-conquer strategy with a dedicated two-stage network architecture. Mosaicing artifacts have also been extensively studied. Early optimization-based methods \cite{morimatsu2021monochrome, qiu2021linear, lu2024hybrid} incorporated handcrafted priors within numerical reconstruction frameworks. More recently, learning-based demosaicing approaches have emerged, leveraging neural networks \cite{zeng2019end, nguyen2022two, guo2024attention, zhou2025pidsr} and dictionary learning techniques \cite{wen2021sparse, zhang2021polarization, luo2024learning}. 

Despite their effectiveness within specific settings, these methods are predominantly designed for individual degradation types, with architectures closely coupled to the underlying corruption model. When adapted to other degradation scenarios, their performance often degrades, suggesting limited architectural versatility. This task-specific structural design paradigm contrasts with the goal of establishing a unified polarization-aware backbone applicable across diverse degradation conditions.

\subsection{Architectural Unification for Conventional RGB Imaging}
To the best of our knowledge, no existing polarimetric imaging method achieves architectural unification across multiple degradations. In contrast, substantial efforts have been devoted to unified restoration frameworks for conventional RGB imaging. 

Early CNN-based approaches such as MPRNet \cite{zamir2021multi} adopt multi-stage progressive designs to handle various restoration tasks within a shared architecture. The field has since shifted toward attention-based models to better capture long-range dependencies. Representative Transformer-based methods, including SwinIR \cite{liang2021swinir}, Restormer \cite{zamir2022restormer}, and Uformer \cite{wang2022uformer}, demonstrate strong cross-task performance within unified frameworks. NAFNet \cite{chen2022simple} further shows that carefully designed streamlined architectures can achieve competitive or even superior results with improved efficiency. Beyond deterministic restoration networks, generative approaches such as DiffIR \cite{xia2023diffir} leverage diffusion models to provide strong priors for diverse degradation scenarios. More recently, state space model (SSM)-based methods, exemplified by VmambaIR \cite{shi2025vmambair}, establish unified restoration frameworks with global receptive fields and linear complexity, further advancing architectural unification in RGB imaging.

Despite their effectiveness in RGB imaging, these architectures do not explicitly model the physical coupling among polarized measurements. In polarimetric imaging, the captured intensities are governed by the Stokes representation and are intrinsically linked through polarization physics. Directly applying RGB-oriented restoration frameworks may therefore lead to physically inconsistent Stokes estimation and unreliable DoP/AoP recovery.

\section{Background}
\subsection{Polarized Images and Polarimetric Parameters}
A polarized measurement $\mathbf{I}_\alpha$ is obtained by placing a linear polarizer with orientation angle $\alpha$ in front of an image sensor. Under the linear polarization model and assuming a linear camera response, the captured intensity follows Malus’ law \cite{hecht2012optics}:
\begin{equation}
    \label{eq: Malus1}
    \mathbf{I}_\alpha = \frac{\mathbf{I}}{2} \cdot (1 - \mathbf{p} \cdot \cos(2(\alpha-\bm{\theta}))),
\end{equation}
where $\mathbf{I}$ denotes the total intensity (TI), corresponding to the image captured without a polarizer. The variables $\mathbf{p} \in [0, 1]$ and $\bm{\theta} \in [0, \pi]$ represent the Degree of Polarization (DoP) and Angle of Polarization (AoP) of the incident light, respectively\footnote{An alternative but equivalent formulation $\mathbf{I}_\alpha = \mathbf{I} \cdot (1 + \mathbf{p} \cdot \cos(2(\bm{\theta} - \alpha))) / 2$ is adopted in some prior works \cite{kalra2020deep, peters2023pcon}, reflecting a different sign convention. We follow the formulation used in \cite{zhou2023polarization, zhou2025learning}. Further discussion on convention differences can be found in \cite{zhou2025polarimetric}.}. For clarity, we restrict our discussion to linear polarization, which is consistent with most practical polarization cameras that employ linear micro-polarizer arrays. To make the linear structure explicit, \Eref{eq: Malus1} can be rewritten in an inner-product form:
\begin{equation}
    \label{eq: Malus2}
    \mathbf{I}_\alpha = \left\langle \left[ \frac{1}{2} \quad \frac{-\cos(2\alpha)}{2} \quad \frac{-\sin(2\alpha)}{2} \right], \vec{\mathbf{S}} \right\rangle,
\end{equation}
where
\begin{equation}
    \label{eq: StokesParameters1}
    \vec{\mathbf{S}}=[\mathbf{S}_0 \quad \mathbf{S}_1 \quad \mathbf{S}_2]^{\top} = [\mathbf{I} \quad \mathbf{I} \cdot \mathbf{p} \cdot \cos(2\bm{\theta}) \quad \mathbf{I} \cdot \mathbf{p} \cdot \sin(2\bm{\theta})]^{\top}
\end{equation}
denotes Stokes parameters under linear polarization \cite{konnen1985polarized}. 

Given the Stokes parameters $\mathbf{S}_{0,1,2}$, the DoP $\mathbf{p}$ and AoP $\bm{\theta}$ can be recovered as:
\begin{equation}
    \label{eq: DoPAoP}
    \mathbf{p}=\frac{\sqrt{\mathbf{S}_1^2 + \mathbf{S}_2^2}}{\mathbf{S}_0} \quad\text{and}\quad \bm{\theta}=\frac{1}{2}\arctan(\frac{\mathbf{S}_2}{\mathbf{S}_1}).
\end{equation}

\subsection{Polarimetric Imaging}
Polarimetric imaging aims to estimate the polarimetric parameters (TI $\mathbf{I}$, DoP $\mathbf{p}$, and AoP $\bm{\theta}$) through recovery of the Stokes parameters $\mathbf{S}_{0,1,2}$. According to \Eref{eq: Malus2}, estimating $\mathbf{S}_{0,1,2}$ requires solving a linear system constructed from measurements captured at different polarizer orientations, implying that at least three distinct polarization angles are necessary. While multiple measurements can be acquired sequentially using a rotating linear polarizer, modern polarization cameras employ micro-polarizer arrays to capture four polarization orientations simultaneously. Specifically, images $\mathbf{I}_{\alpha_{1,2,3,4}}$ are recorded at angles $\alpha_{1,2,3,4}=0^{\circ}, 45^{\circ}, 90^{\circ}, 135^{\circ}$ within a single exposure. Substituting these angles into \Eref{eq: Malus1} yields closed-form solutions for the Stokes parameters:
\begin{equation}
    \label{eq: StokesParameters2}
    \begin{dcases}
        \mathbf{S}_0 = 2\bar{\mathbf{I}}_{\alpha_i} = \mathbf{I}_{\alpha_1} + \mathbf{I}_{\alpha_3} = \mathbf{I}_{\alpha_2} + \mathbf{I}_{\alpha_4}\\
        \mathbf{S}_1 = \mathbf{I}_{\alpha_3} - \mathbf{I}_{\alpha_1}\\
        \mathbf{S}_2 = \mathbf{I}_{\alpha_4} - \mathbf{I}_{\alpha_2}
    \end{dcases}
    ,
\end{equation}
where $\bar{\mathbf{I}}_{\alpha_i}$ denotes the average polarized image, which can be written as:
\begin{equation}
    \bar{\mathbf{I}}_{\alpha_i} = \frac{1}{4} \sum_{i=1}^4 \mathbf{I}_{\alpha_i}.
\end{equation}

Although the recovery of the Stokes parameters can be formulated as a linear inversion problem, the mapping from Stokes space to DoP and AoP is inherently nonlinear (see \Eref{eq: DoPAoP}, which involves division and inverse trigonometric operations). As a result, small perturbations in the measured intensities may induce disproportionately large deviations in the estimated DoP and AoP. This nonlinear sensitivity poses significant challenges for robust polarimetric imaging under degraded conditions. An empirical illustration of this effect can be observed in \Fref{fig: Teaser} (e), where degradations such as low-light noise, motion blur, and mosaicing artifacts lead to noticeable distortions in the DoP and AoP.

\begin{figure*}[t]
    \centering
    \includegraphics[width=1.0\linewidth]{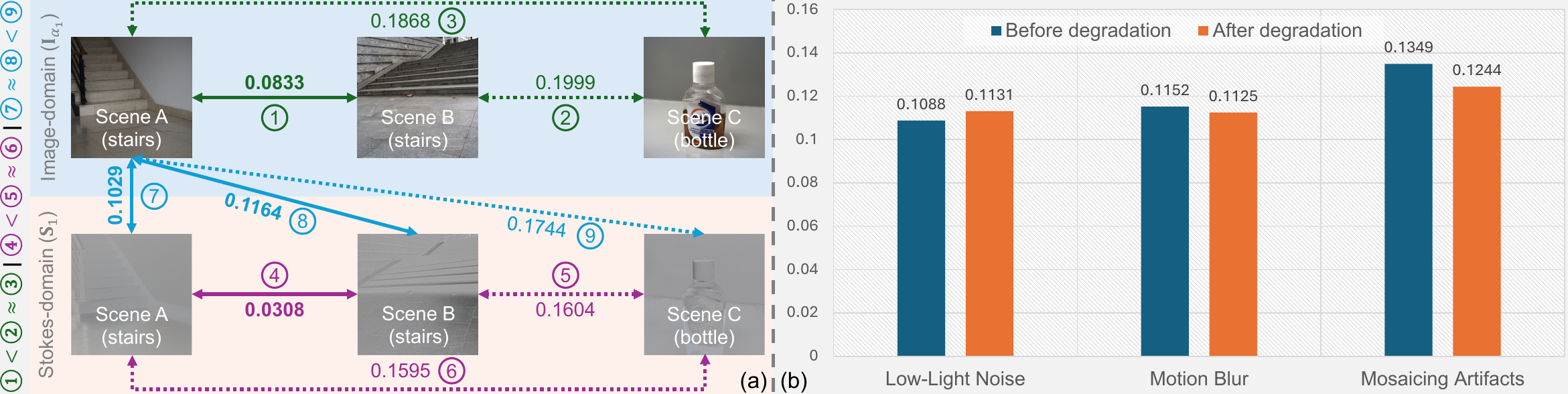}
    \caption{Empirical validation of cross-domain semantic consistency. (a) A sanity-check experiment demonstrating that CLIP-based embeddings preserve coarse semantic relationships in both the image (\ie, $\mathbf{I}_{\alpha_1}$) and Stokes (\ie, $\mathbf{S}_1$) domains. The measured semantic distances ($d_{\text{sem}}$) between scenes of the same class are consistently smaller than those between different classes. (b) Quantitative analysis across three distinct degradation tasks, including low-light noise (using the PLIE dataset \cite{zhou2023polarization}), motion blur (using the PolDeblur dataset \cite{zhou2025learning}), and mosaicing artifacts (using the PIDSR dataset \cite{zhou2025pidsr}). The results indicate that the expected cross-domain semantic distance ($\mathbb{E}[d_{\text{sem}}]$) remains remarkably stable before and after degradation, with a relative deviation consistently below $10\%$.}
    \label{fig: Verification}
\end{figure*}

\section{Design Motivation and Problem Formulation}
\subsection{Structural Analysis of Existing Paradigms}
In real-world environments, captured polarized images are frequently degraded by various challenging factors. First, low-light conditions severely reduce the signal-to-noise ratio (SNR) due to limited photon counts \cite{hu2020iplnet, xu2022colorpolarnet, zhou2023polarization}. Furthermore, motion blur is nearly inevitable on moving platforms, as the inherent light attenuation caused by polarizers necessitates longer exposure times \cite{zhou2025learning}. Additionally, hardware-specific issues arise in division-of-focal-plane (DoFP) polarization cameras, where the requisite demosaicing process (typically bilinear interpolation) often introduces severe mosaicing artifacts \cite{zhou2025pidsr}. Consequently, the recovered polarimetric parameters can deviate significantly from their true physical values, leading to distorted scene interpretations and compromised performance in downstream applications.

To mitigate these severe degradations, various computational restoration methods have been proposed; however, their architectural paradigms often exhibit inherent limitations. Formally, the physical degradation process in real-world polarimetric imaging can be generalized as:
\begin{equation}
    \mathbf{I}^{*}_{\alpha_{1,2,3,4}} = \mathcal{D}(\mathbf{I}_{\alpha_{1,2,3,4}}) + \mathbf{N}_{\alpha_{1,2,3,4}},
\end{equation}
where $\mathbf{I}_{\alpha_{1,2,3,4}}$ and $\mathbf{I}^{*}_{\alpha_{1,2,3,4}}$ denote the latent high-quality polarized images and their degraded observations, respectively.\footnote{Throughout this paper, the superscript $*$ indicates a degraded variable.} The operator $\mathcal{D}(\cdot)$ represents specific degradation functions (\eg, motion blur kernel or mosaicing sampling), and $\mathbf{N}_{\alpha_{1,2,3,4}}$ indicates measurement noise. The fundamental problem of polarimetric restoration is an ill-posed inverse problem: reconstructing the clean polarization states $\mathbf{I}_{\alpha_{1,2,3,4}}$ and simultaneously deriving the accurate Stokes parameters $\mathbf{S}_{0,1,2}$. Since existing methods are typically tailored to individual forms of $\mathcal{D}(\cdot)$, it remains unclear whether a structurally consistent framework can be developed without redesigning task-specific architectures.

A straightforward strategy is to perform single-domain restoration using a direct mapping:
\begin{equation}
    \mathbf{I}^{*}_{\alpha_{1,2,3,4}} \xrightarrow{\mathcal{F}_{\text{image}}} \mathbf{I}_{\alpha_{1,2,3,4}} \quad \text{or} \quad \mathbf{S}^{*}_{0,1,2} \xrightarrow{\mathcal{F}_{\text{Stokes}}} \mathbf{S}_{0,1,2},
\end{equation}
as illustrated in \Fref{fig: Teaser} (c). Such single-stage designs are structurally simple but operate within a single representation domain, limiting their ability to exploit cross-domain physical correlations. An alternative is to incorporate cross-domain knowledge through staged designs. For example, one may guide image-domain restoration with Stokes guidance:
\begin{equation}
    \label{eq: StokesGuidancePipeline}
    \mathbf{I}^{*}_{\alpha_{1,2,3,4}} \xrightarrow{\mathcal{F}_1(\cdot,\mathbf{S}^{*}_{0,1,2})} \xrightarrow{\mathcal{F}_2(\cdot,\mathbf{S}^{*}_{0,1,2})} \cdots \xrightarrow{\mathcal{F}_K(\cdot,\mathbf{S}^{*}_{0,1,2})} \mathbf{I}_{\alpha_{1,2,3,4}},
\end{equation}
or perform sequential restoration with domain switching:
\begin{equation}
    \mathbf{I}^{*}_{\alpha_{1,2,3,4}} \xrightarrow{\mathcal{F}_1} \mathbf{I}^{\#}_{\alpha_{1,2,3,4}} \xrightarrow{\text{domain switching}} \mathbf{S}^{\#}_{0,1,2} \xrightarrow{\mathcal{F}_2} \mathbf{S}_{0,1,2},
\end{equation}
as shown in \Fref{fig: Teaser} (a) and (b), respectively. While these approaches leverage cross-domain representations, they rely on multi-stage architectures, which introduce error accumulation. 

To overcome the above limitations, we conceptualize a single-stage multi-domain paradigm that performs joint restoration:
\begin{equation}
    \{\mathbf{I}^{*}_{\alpha_{1,2,3,4}}, \mathbf{S}^{*}_{0,1,2}\} \xrightarrow{\mathcal{F}_{\text{joint}}} \{\mathbf{I}_{\alpha_{1,2,3,4}}, \mathbf{S}_{0,1,2}\},
\end{equation}
as conceptually illustrated in \Fref{fig: Teaser} (d). Unlike the staged approach in \Eref{eq: StokesGuidancePipeline}, where Stokes parameters merely serve as external auxiliary guidance fed into cascaded sub-networks, our paradigm treats both domains as co-equal inputs. By simultaneously ingesting and intrinsically fusing these dual-domain representations within a unified network $\mathcal{F}_{\text{joint}}$, we can theoretically mitigate multi-stage error accumulation. This leads to a fundamental question: \textit{Can we effectively design such a structurally unified framework that preserves single-stage simplicity while intrinsically modeling cross-domain polarization physics?}

\subsection{Empirical Evidence of Cross-Domain Consistency}
To answer the above question affirmatively, it is essential to first establish whether a robust, underlying relationship exists between the image and Stokes domains, particularly when subjected to severe real-world degradations. If both domains are perturbed synergistically, it would strongly justify processing them jointly within a unified architecture. Conversely, if degradations affected the two domains independently, we would expect their semantic correspondence to deteriorate significantly. To verify this hypothesis, we investigate the cross-domain semantic consistency between polarized images and their corresponding Stokes parameters. To measure this high-level semantic distance, we employ the Contrastive Language-Image Pre-training (CLIP) model \cite{radford2021learning}. Let $\Phi_{\text{CLIP}}(\cdot)$ denote the CLIP visual encoder. We formally define the cross-domain semantic distance $d_{\text{sem}}$ between an image-domain representation $\mathbf{r}_i$ and a Stokes-domain representation $\mathbf{r}_s$ as the Mean Squared Error (MSE) of their encoded semantic feature maps \cite{chefer2022image}:
\begin{equation}
    d_{\text{sem}}(\mathbf{r}_i, \mathbf{r}_s) = \text{MSE}\left(\Phi_{\text{CLIP}}(\mathbf{r}_i), \Phi_{\text{CLIP}}(\mathbf{r}_s)\right).
\end{equation}
While CLIP is pre-trained on natural RGB images, we argue it is highly applicable to Stokes parameters. As indicated by \Eref{eq: StokesParameters2}, $\mathbf{S}_{1,2}$ can be interpreted as differential signals computed from polarized images. Such differential signals preserve essential scene structure and spatial layout, making them compatible with generic visual encoders at a coarse semantic level. Prior studies (\eg, CLIPasso \cite{vinker2022clipasso}) have further demonstrated that CLIP can extract meaningful semantics from non-photorealistic inputs like sketches.

To empirically validate CLIP's applicability to Stokes data, we design a sanity-check experiment using three scenes (A, B, and C), where A and B share the same semantic class (stairs) and C is structurally distinct (bottle). Without loss of generality, let $\mathbf{I}_{\alpha_1}^{(k)}$ and $\mathbf{S}_1^{(k)}$ denote the selected image and Stokes representations of Scene $k \in \{A, B, C\}$. As shown in \Fref{fig: Verification} (a), the semantic distances between same-class scenes are consistently smaller than those between different classes in both domains:
\begin{equation}
    d_{\text{sem}}(\mathbf{D}^{(A)}, \mathbf{D}^{(B)}) < d_{\text{sem}}(\mathbf{D}^{(A)}, \mathbf{D}^{(C)}) \approx d_{\text{sem}}(\mathbf{D}^{(B)}, \mathbf{D}^{(C)}),
\end{equation}
where $\mathbf{D}^{(k)} \in \{\mathbf{I}_{\alpha_1}^{(k)}, \mathbf{S}_1^{(k)}\}$, corresponding to \ding{172} $<$ \ding{173} $\approx$ \ding{174} and \ding{175} $<$ \ding{176} $\approx$ \ding{177} in the figure.\footnote{In our implementation, both polarized images and Stokes parameters are linearly normalized to $[0, 1]$, converted to 8-bit images, and processed using CLIP’s standard preprocessing pipeline to ensure numerical stability.} More importantly, this relative ordering holds perfectly in the cross-domain setting: the image of Scene A is semantically closer to the Stokes representations of Scenes A and B than to that of Scene C:
\begin{equation}
    d_{\text{sem}}(\mathbf{I}_{\alpha_1}^{(A)}, \mathbf{S}_1^{(A)}) \approx d_{\text{sem}}(\mathbf{I}_{\alpha_1}^{(A)}, \mathbf{S}_1^{(B)}) < d_{\text{sem}}(\mathbf{I}_{\alpha_1}^{(A)}, \mathbf{S}_1^{(C)}),
\end{equation}
which perfectly aligns with the visual order \ding{178} $\approx$ \ding{179} $<$ \ding{180}. This sanity check rigorously confirms that CLIP embeddings can reliably capture cross-domain semantic correspondences.

Building upon this validated measurement tool, we investigate how severe image degradations affect this cross-domain consistency. We conduct a quantitative analysis under three representative degradation types: low-light noise (using the PLIE dataset \cite{zhou2023polarization}), motion blur (using the PolDeblur dataset \cite{zhou2025learning}), and mosaicing artifacts (using the PIDSR dataset \cite{zhou2025pidsr}). Specifically, we measure the expected semantic distance over the entire dataset between the clean pairs, denoted as $\mathbb{E}[d_{\text{sem}}(\mathbf{I}_{\alpha_1}, \mathbf{S}_1)]$, and their corresponding degraded pairs, denoted as $\mathbb{E}[d_{\text{sem}}(\mathbf{I}^*_{\alpha_1}, \mathbf{S}^*_1)]$, where $\mathbb{E}[\cdot]$ represents the mean computation across all scenes. As illustrated in \Fref{fig: Verification} (b), across all three diverse degradation types, the post-degradation cross-domain distance remains remarkably stable compared to the pre-degradation baseline:
\begin{equation}
    \mathbb{E}\left[d_{\text{sem}}(\mathbf{I}^*_{\alpha_1}, \mathbf{S}^*_1)\right] \approx \mathbb{E}\left[d_{\text{sem}}(\mathbf{I}_{\alpha_1}, \mathbf{S}_1)\right],
\end{equation}
with the relative deviation consistently below 10\%. This empirical evidence reveals that degradations perturb both domains synergistically while preserving their intrinsic semantic linkage. Consequently, this strongly motivates the design of a unified cross-domain architecture that jointly processes image- and Stokes-domain representations, rather than treating them in isolation.

\begin{figure*}[t]
    \centering
    \includegraphics[width=1.0\linewidth]{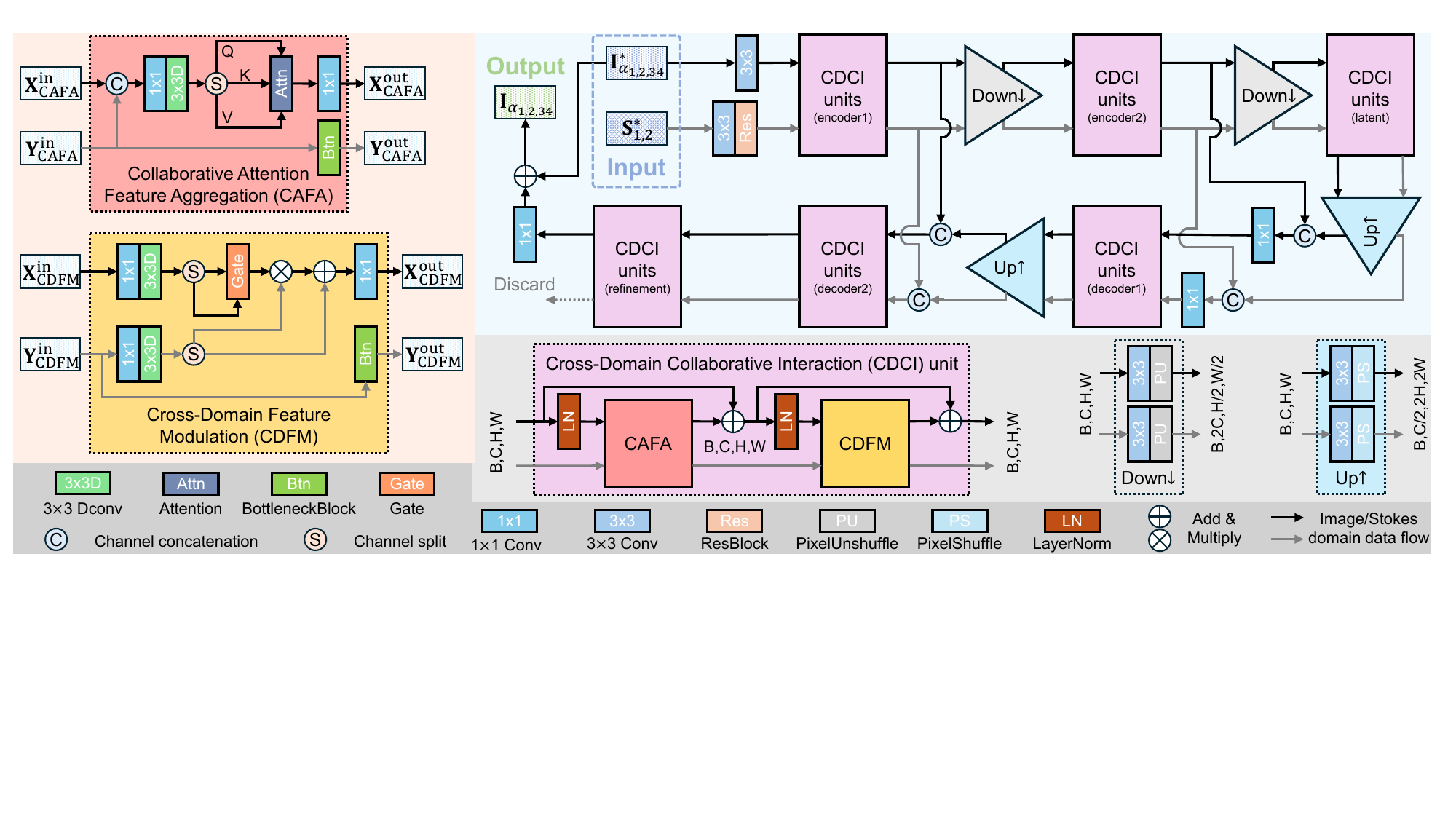}
    \caption{Overview of our single-stage multi-domain architecture. The framework is built upon a dual-branch U-shaped backbone that jointly processes representations from both the image and Stokes domains. To facilitate the information flow between these domains, the core components, Cross-Domain Collaborative Interaction (CDCI) units, inherently integrate Collaborative Attention Feature Aggregation (CAFA) and Cross-Domain Feature Modulation (CDFM).}
    \label{fig: Network}
\end{figure*}

\section{Method}
\subsection{Overall Architecture}
As illustrated in \Fref{fig: Network} (top right), our framework adopts a single-stage multi-domain architecture structured upon a U-shaped hierarchical backbone \cite{ronneberger2015u}. Unlike single-domain methods, our network explicitly accepts inputs from two distinct physical domains: the degraded polarized images $\mathbf{I}^{*}_{\alpha_{1,2,3,4}}$ and their corresponding degraded Stokes parameters $\mathbf{S}^{*}_{1,2}$. It is worth noting that $\mathbf{S}^*_0$ is explicitly excluded from the Stokes input branch, as it is mathematically equivalent to the total intensity $\mathbf{I}$ (according to \Eref{eq: StokesParameters1}), which is inherently embedded within the image domain. To map the raw inputs into high-dimensional feature spaces, we employ two domain-specific shallow feature extractors. A $3 \times 3$ convolutional layer is applied to the polarized images to extract initial image-domain features $\mathbf{X}_0$. Concurrently, a $3 \times 3$ convolution followed by a ResBlock \cite{he2016deep} is applied to the Stokes parameters to extract the initial Stokes-domain features $\mathbf{Y}_0$.

Following the U-shaped topology, these multi-domain features are progressively processed through a symmetric dual-branch encoder-decoder structure. The encoder pathway consists of two hierarchical levels of CDCI units, each followed by a downsampling block (strided convolutions) to aggregate spatial context and increase the receptive field. At the bottleneck, a latent CDCI module processes the highly abstracted features. The decoder pathway mirrors the encoder, utilizing upsampling blocks (pixel-shuffle operations \cite{shi2016real}) followed by CDCI units to progressively recover the spatial resolution. To mitigate the loss of high-frequency spatial details caused by downsampling, skip connections are established to fuse the intermediate features from the encoder directly to their corresponding decoder levels.

Finally, a refinement module (composed of CDCI units) operates at the original resolution to further restore fine-grained polarimetric textures. Instead of directly regressing the target images, the network predicts the residuals between the degraded inputs and the clean targets, which significantly eases the optimization process. Denoting the entire network mapping as $\mathcal{F}_{\text{net}}$, the overall reconstruction process is formulated as:
\begin{equation}
    \mathbf{I}_{\alpha_{1,2,3,4}} = \mathbf{I}^{*}_{\alpha_{1,2,3,4}} + \mathcal{F}_{\text{net}}(\mathbf{I}^{*}_{\alpha_{1,2,3,4}}, \mathbf{S}^{*}_{1,2}),
\end{equation}
where $\mathbf{I}_{\alpha_{1,2,3,4}}$ represents the restored high-quality polarized images, from which the accurate physical Stokes parameters can be analytically derived.

\subsection{Cross-Domain Collaborative Interaction (CDCI) Unit}
While the macro-architecture is multi-domain, the fundamental challenge lies in how to effectively fuse and modulate the information \textit{across} these distinct representations. To this end, we introduce the CDCI unit as the core building block across all modules (encoder, latent, decoder, and refinement) of our backbone.

As shown in \Fref{fig: Network} (bottom), each CDCI unit is designed with a dual-branch topology to maintain the independent physical significance of both domains while facilitating their cross-domain interaction. Let $\{\mathbf{X}_{\text{in}}, \mathbf{Y}_{\text{in}}\}$, $\{\mathbf{X}_{\text{mid}}, \mathbf{Y}_{\text{mid}}\}$, and $\{\mathbf{X}_{\text{out}}, \mathbf{Y}_{\text{out}}\}$ denote the input, intermediate, and output features for the image and Stokes branches, respectively. The unit sequentially executes two highly coupled operations: Collaborative Attention Feature Aggregation (CAFA) and Cross-Domain Feature Modulation (CDFM). Mathematically, the forward pass through a CDCI unit is governed by the following formulations:
\begin{equation}
    \label{eq: CDCI}
    \begin{dcases}
        \mathbf{X}_{\text{mid}}, \mathbf{Y}_{\text{mid}} = \text{CAFA}(\text{LN}(\mathbf{X}_{\text{in}}), \mathbf{Y}_{\text{in}}) + (\mathbf{X}_{\text{in}}, \mathbf{0})\\
        \mathbf{X}_{\text{out}}, \mathbf{Y}_{\text{out}} = \text{CDFM}(\text{LN}(\mathbf{X}_{\text{mid}}), \mathbf{Y}_{\text{mid}}) + (\mathbf{X}_{\text{mid}}, \mathbf{0})
    \end{dcases}
    ,
\end{equation}
where $\text{LN}(\cdot)$ represents Layer Normalization \cite{ba2016layer}.

Crucially, as indicated by the residual additions in \Eref{eq: CDCI}, the image-domain branch employs dense residual connections, whereas the Stokes-domain branch operates in a straightforward feed-forward manner. This asymmetric design choice is physically motivated: the image-domain signals encapsulate rich spatial details (\eg, textures), for which residual learning is vital to ensure stable gradient flow and fine-grained refinement. Conversely, the Stokes parameters ($\mathbf{S}_{1,2}$) are fundamentally ``differential signals'' (see \Eref{eq: StokesParameters2}) that inherently encode gradient-like structural cues; thus, they do not require the complexity of continuous residual refinement, allowing for a more streamlined feature propagation.

\subsection{Collaborative Attention Feature Aggregation (CAFA)}
The CAFA module, illustrated in \Fref{fig: Network} (top left), is specifically tailored to aggregate complementary information across the two domains. While the Stokes features guide the structural alignment, the image features provide the necessary textural context.

For the image-domain branch, we leverage a cross-channel self-attention mechanism \cite{zamir2022restormer, wang2022uformer} to model global context. First, we concatenate the dual-branch inputs $\mathbf{X}^{\text{in}}_{\text{CAFA}}$ and $\mathbf{Y}^{\text{in}}_{\text{CAFA}}$ along the channel dimension. A $1 \times 1$ point-wise convolution followed by a $3 \times 3$ depth-wise convolution ($\text{DConv}$) is then applied to efficiently encode spatially local representations:
\begin{equation}
    \mathbf{H} = \text{DConv}_{3\times3}(\text{Conv}_{1\times1}([\mathbf{X}^{\text{in}}_{\text{CAFA}}, \mathbf{Y}^{\text{in}}_{\text{CAFA}}])),
\end{equation}
where $[\cdot, \cdot]$ denotes concatenation. The aggregated feature $\mathbf{H}$ is subsequently split into three distinct projections: Query ($\mathbf{Q}$), Key ($\mathbf{K}$), and Value ($\mathbf{V}$). To compute the self-attention across the channel dimension, these tensors are reshaped such that $\mathbf{Q} \in \mathbb{R}^{HW \times C}$, $\mathbf{K} \in \mathbb{R}^{C \times HW}$, and $\mathbf{V} \in \mathbb{R}^{HW \times C}$. The cross-channel attention is then formulated as:
\begin{equation}
    \text{Attention}(\mathbf{Q}, \mathbf{K}, \mathbf{V}) = \mathbf{V} \otimes \text{Softmax}(\mathbf{K} \otimes \mathbf{Q} / \tau),
\end{equation}
where $\otimes$ denotes the matrix multiplication, and $\tau$ is a learnable temperature parameter designed to dynamically scale the attention map before applying the Softmax function. This channel-wise attention efficiently captures long-range dependencies and global context without incurring the quadratic spatial complexity of standard vision Transformers \cite{vaswani2017attention, dosovitskiy2020image}. Finally, a $1 \times 1$ convolution refines the attention output, and a residual connection is added to produce $\mathbf{X}^{\text{out}}_{\text{CAFA}}$.

For the Stokes-domain branch, we eschew complex attention modeling in favor of a robust, convolution-based BottleneckBlock \cite{he2016deep}, enhanced with Instance Normalization \cite{ulyanov2016instance} and \texttt{ReLU} activations. This design effectively extracts the local structural gradients inherent in the differential Stokes signals $\mathbf{Y}^{\text{in}}_{\text{CAFA}}$ to yield $\mathbf{Y}^{\text{out}}_{\text{CAFA}}$.

\subsection{Cross-Domain Feature Modulation (CDFM)}
Following cross-domain feature aggregation, the CDFM module (\Fref{fig: Network}, middle left) is introduced to perform precise, spatially-varying feature modulation. This module enables the Stokes-domain structural priors to dynamically guide and modulate the image-domain feature restoration.

In the image-domain branch, the input features $\mathbf{X}^{\text{in}}_{\text{CDFM}}$ undergo a $1 \times 1$ convolution and a $3 \times 3$ depth-wise convolution. The resulting tensor is split along the channel dimension into two distinct flows: an input flow $\mathbf{F}^{\text{x}}_{\text{i}}$ and a gating flow $\mathbf{F}^{\text{x}}_{\text{g}}$. The gating flow is activated by a \texttt{GELU} function \cite{hendrycks2016gaussian} and performs element-wise multiplication with the input flow, serving as an intrinsic feature filter. Crucially, the modulation parameters are derived from the Stokes-domain branch. The input Stokes features $\mathbf{Y}^{\text{in}}_{\text{CDFM}}$ are processed through an identical sequence of convolutions and split into a scaling multiplier $\mathbf{F}^{\text{y}}_{\text{m}}$ and a shifting bias $\mathbf{F}^{\text{y}}_{\text{b}}$. These Stokes-derived parameters apply a linear affine transformation to the gated image features. Denoting the modulated intermediate result as $\mathbf{F}^{\text{x}}_{\text{mod}}$, the cross-domain modulation process is rigorously formulated as:
\begin{equation}
    \mathbf{F}^{\text{x}}_{\text{mod}} = (\texttt{GeLU}(\mathbf{F}^{\text{x}}_{\text{g}}) \cdot \mathbf{F}^{\text{x}}_{\text{i}}) \cdot \mathbf{F}^{\text{y}}_{\text{m}} + \mathbf{F}^{\text{y}}_{\text{b}}.
\end{equation}
A final $1 \times 1$ convolution is applied to $\mathbf{F}^{\text{x}}_{\text{mod}}$ to produce the modulated output $\mathbf{X}^{\text{out}}_{\text{CDFM}}$.

Parallel to the CAFA design, the Stokes-domain branch in CDFM processes $\mathbf{Y}^{\text{in}}_{\text{CDFM}}$ using an enhanced BottleneckBlock to output $\mathbf{Y}^{\text{out}}_{\text{CDFM}}$. This explicit cross-domain modulation mechanism ensures that the reconstructed polarized images strictly adhere to the underlying physical structure dictated by the Stokes parameters, thereby offering significantly richer representational capacity and physical fidelity compared to isolated single-domain processing.

\section{Implementation Details}
\subsection{Objective Functions}
To effectively train our single-stage multi-domain architecture, we design a comprehensive objective function that supervises the network in both the image and Stokes domains. The overall loss function $\mathcal{L}$ is formulated as a weighted sum:
\begin{equation}
    \mathcal{L} = \mathcal{L}_{\text{i}} + \lambda \mathcal{L}_{\text{s}},
\end{equation}
where $\mathcal{L}_{\text{i}}$, and $\mathcal{L}_{\text{s}}$ denote the image-domain loss and Stokes-domain loss, respectively. $\lambda$ is a balancing hyperparameter empirically set to $10$ to ensure the Stokes-domain gradients are adequately scaled relative to the image domain.

\textbf{Image-domain loss.} The image-domain loss $\mathcal{L}_{\text{i}}$ aims to restore the textural fidelity and intensity accuracy of the polarized images. Let the subscript $\text{gt}$ denote the ground-truth signals. $\mathcal{L}_{\text{i}}$ is defined as:
\begin{equation}
    \mathcal{L}_{\text{i}} = \sum_{k=1}^4 \Big( \mathcal{L}_1(\mathbf{I}_{\alpha_k}, \mathbf{I}^{\text{gt}}_{\alpha_k}) + \lambda_1 \mathcal{L}_{\text{p}}(\mathbf{I}_{\alpha_k}, \mathbf{I}^{\text{gt}}_{\alpha_k}) \Big) + \lambda_2 \mathcal{R}_{\text{i}}(\mathbf{I}_{\alpha_{1,2,3,4}}),
\end{equation}
where $\mathcal{L}_1$ is the standard $\ell_1$ pixel-wise loss, and $\mathcal{L}_{\text{p}}$ represents the perceptual loss. The term $\mathcal{R}_{\text{i}}$ is a physics-based regularization aiming to enforce \Eref{eq: StokesParameters2}:
\begin{equation}
    \mathcal{R}_{\text{i}}(\mathbf{I}_{\alpha_{1,2,3,4}}) = \ell_1(\mathbf{I}_{\alpha_1} + \mathbf{I}_{\alpha_3}, \mathbf{I}_{\alpha_2} + \mathbf{I}_{\alpha_4}).
\end{equation}
The weighting coefficients are set to $\lambda_1 = 0.01$ and $\lambda_2 = 1$.

\textbf{Stokes-domain loss.} To ensure the accuracy of the derived polarimetric parameters, we impose explicit supervision on the Stokes domain. The Stokes-domain loss $\mathcal{L}_{\text{s}}$ is defined as:
\begin{equation}
    \mathcal{L}_{\text{s}} = \sum_{k=1}^2 \mathcal{L}_1(\mathbf{S}_k, \mathbf{S}^{\text{gt}}_k) + \mathcal{R}_{\text{s}}(\mathbf{S}_{1,2}, \mathbf{S}^{\text{gt}}_{1,2}),
\end{equation}
where $\mathcal{R}_{\text{s}}$ is the cross-consistency regularization term inherently designed to optimize the AoP. Since AoP is computed via an arctangent function (see \Eref{eq: DoPAoP}), directly applying a loss on AoP can cause severe numerical instability (\eg, singularity when $\mathbf{S}_1 \to 0$ or phase wrapping issues). To circumvent this, we mathematically reformulate the angular constraint. By ensuring the ratio $\mathbf{S}_1 / \mathbf{S}_2 \approx \mathbf{S}^{\text{gt}}_1 / \mathbf{S}^{\text{gt}}_2$, we can penalize the angular deviation through a stable cross-product formulation:
\begin{equation}
    \mathcal{R}_{\text{s}}(\mathbf{S}_{1,2}, \mathbf{S}^{\text{gt}}_{1,2}) = \ell_1(\mathbf{S}_1 \cdot \mathbf{S}^{\text{gt}}_2, \mathbf{S}_2 \cdot \mathbf{S}^{\text{gt}}_1).
\end{equation}
This formulation promotes structural alignment between the reconstructed and ground-truth Stokes vectors, ensuring highly accurate AoP recovery while avoiding the numerical pitfalls of inverse trigonometric functions.

\begin{figure*}[t]
    \centering
    \includegraphics[width=1.0\linewidth]{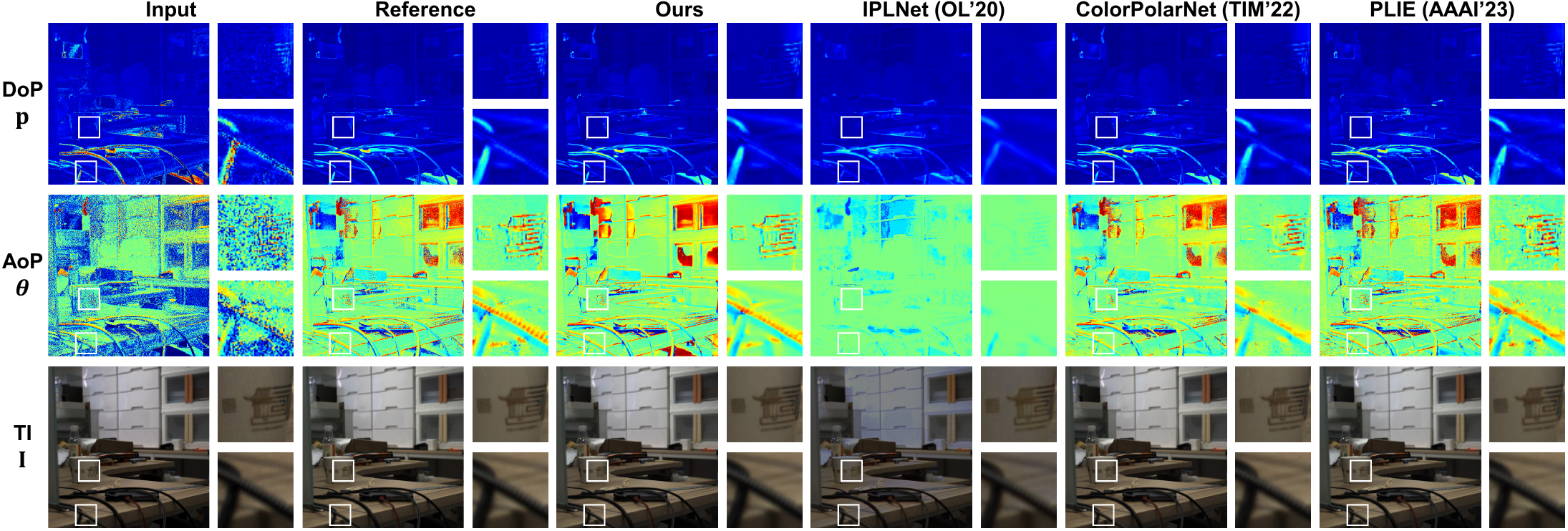}
    \caption{Qualitative comparisons on the real-world PLIE dataset \cite{zhou2023polarization} for polarimetric recovery under low-light noise. Note that the input TI ($\mathbf{I}$) has been digitally amplified for visualization purposes, as the original low-light captures are nearly entirely dark. Our method effectively suppresses noise while accurately preserving the structural details of DoP and AoP. Please zoom in for better details.}
    \label{fig: LowLight}
\end{figure*}

\begin{table*}[t]
    \caption{Quantitative comparisons on the real-world PLIE dataset \cite{zhou2023polarization} for polarimetric recovery under low-light noise. The best and second-best results are highlighted in \textbf{bold} and \underline{underline}, respectively.}
    \label{tab: LowLight}
    \centering
    \begin{tabular}{lcccccc}
        \toprule
        Method 
        & PSNR-DoP & SSIM-DoP 
        & PSNR-AoP & SSIM-AoP 
        & PSNR-TI  & SSIM-TI \\
        \midrule
        Ours 
        & \textbf{30.61} & \textbf{0.842} 
        & \textbf{16.69} & \textbf{0.446} 
        & \textbf{40.78} & \textbf{0.986} \\
        IPLNet \cite{hu2020iplnet} 
        & 25.32 & 0.715 
        & 16.21 & 0.276 
        & 22.84 & 0.930 \\
        ColorPolarNet \cite{xu2022colorpolarnet} 
        & \underline{28.32} & \underline{0.785} 
        & 16.37 & 0.332 
        & \underline{40.06} & \underline{0.983} \\ 
        PLIE \cite{zhou2023polarization} 
        & 27.15 & 0.765 
        & \underline{16.42} & \underline{0.336} 
        & 39.19 & 0.977 \\
        \bottomrule
    \end{tabular}
\end{table*}

\subsection{Implementation Details}
Our proposed unified architecture is implemented using the PyTorch framework and optimized on a single NVIDIA RTX 4090 GPU. The backbone is configured with six modules: two encoder levels, one latent bottleneck, two decoder levels, and one refinement module. To balance computational efficiency and representational capacity, the number of CDCI units within these modules is empirically set to $[4, 6, 6, 6, 4, 4]$, respectively. Within the CAFA module, the number of attention heads is configured as $[1, 2, 4, 2, 1, 1]$. This progressively increasing number of heads enables the network to capture broader, more complex global contexts at lower spatial resolutions, while maintaining computational efficiency at higher resolutions. Furthermore, the hidden feature channels inside the CDFM module are expanded to twice the input dimensionality, providing sufficient capacity for cross-domain modulation.

During the training phase, we adopt the AdamW optimizer \cite{loshchilov2017decoupled} with default momentum parameters ($\beta_1=0.9, \beta_2=0.999$) and a weight decay of $1 \times 10^{-5}$ to prevent overfitting. The network is trained end-to-end for a total of $3 \times 10^5$ epochs. We initialize the learning rate at $3 \times 10^{-4}$ and gradually decay it to a minimum of $1 \times 10^{-6}$ following the cosine annealing schedule \cite{loshchilov2016sgdr}, which ensures stable convergence. 

\section{Experiments}
To rigorously demonstrate the versatility and superiority of our proposed unified architecture, we conduct extensive evaluations across three distinct and challenging degradation tasks prevalent in real-world polarimetric imaging: (a) low-light noise, (b) motion blur, and (c) mosaicing artifacts. Crucially, aligning with our core philosophy of \textit{architectural unification}, we utilize the exact same network topology and hyperparameter configurations across all tasks. For a fair and task-specific evaluation, we train separate model weights (instantiations) for each degradation type.

\textbf{Baselines and evaluation metrics.} We benchmark our framework primarily against state-of-the-art methods specifically designed for polarimetric imaging. As demonstrated in prior literature, conventional RGB-based restoration models often fail to preserve underlying physical constraints, leading to severe distortions in polarization states. Following standard protocols adopted by prior works \cite{hu2020iplnet, xu2022colorpolarnet, zhou2023polarization, zhou2025learning, zhou2025pidsr}, we evaluate not only the DoP ($\mathbf{p}$) and AoP ($\bm{\theta}$) but also the TI ($\mathbf{I}$). We report the Peak Signal-to-Noise Ratio (PSNR) and Structural Similarity Index (SSIM) to assess both pixel-wise fidelity and structural coherence.

\begin{figure*}[t]
    \centering
    \includegraphics[width=1.0\linewidth]{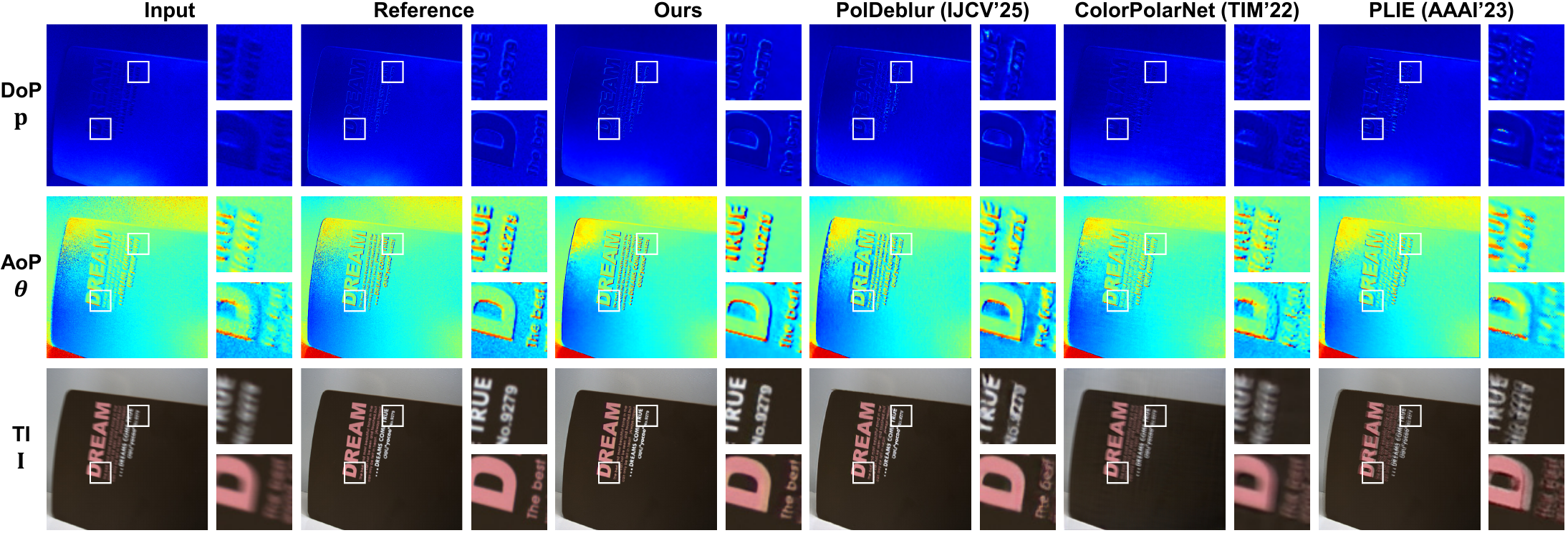}
    \caption{Qualitative comparisons on the synthetic PolDeblur dataset \cite{zhou2025learning} for handling motion blur. Our method effectively restores sharp textures (\eg, the text below the character ``D'') while accurately reconstructing polarimetric states without ringing artifacts. Please zoom in for better details.}
    \label{fig: MotionBlurS}
\end{figure*}

\begin{figure*}[t]
    \centering
    \includegraphics[width=1.0\linewidth]{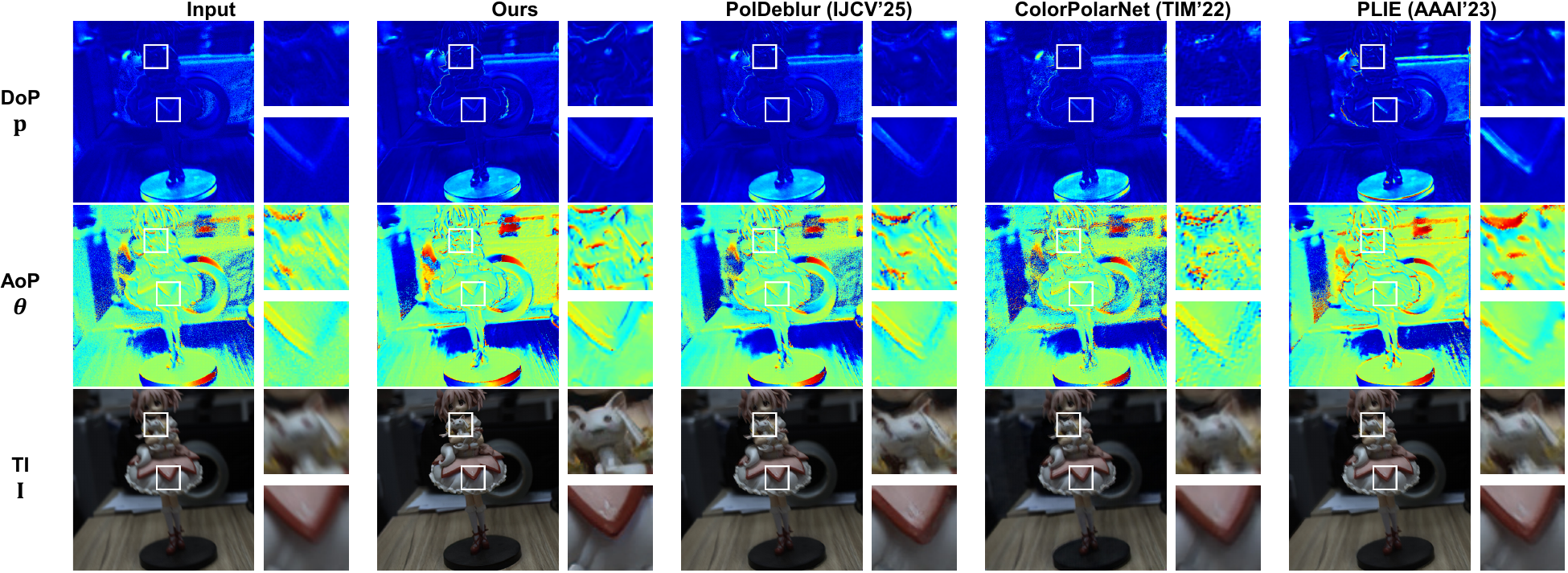}
    \caption{Qualitative evaluations on real-captured degraded polarized images. Despite being trained solely on synthetic pairs, our method demonstrates remarkable generalization capability in removing real-world motion blur and recovering faithful polarization structures. Please zoom in for better details.}
    \label{fig: MotionBlurR}
\end{figure*}

\begin{table*}[t]
    \caption{Quantitative comparisons on the synthetic PolDeblur dataset \cite{zhou2025learning} for handling motion blur. The best and second-best results are highlighted in \textbf{bold} and \underline{underline}, respectively.}
    \label{tab: MotionBlur}
    \centering
    \begin{tabular}{lcccccc}
        \toprule
        Method 
        & PSNR-DoP & SSIM-DoP 
        & PSNR-AoP & SSIM-AoP 
        & PSNR-TI  & SSIM-TI \\
        \midrule
        Ours 
        & \textbf{30.31} & \textbf{0.804} 
        & \textbf{17.92} & \textbf{0.378} 
        & \textbf{34.56} & \textbf{0.932} \\
        PolDeblur \cite{zhou2025learning} 
        & \underline{29.12} & \underline{0.774} 
        & \underline{17.77} & \underline{0.330} 
        & \underline{29.67} & \underline{0.868} \\
        ColorPolarNet \cite{xu2022colorpolarnet} 
        & 28.41 & 0.740 
        & 16.99 & 0.279 
        & 25.09 & 0.812 \\
        PLIE \cite{zhou2023polarization} 
        & 26.32 & 0.744 
        & 17.24 & 0.269 
        & 28.40 & 0.853 \\
        \bottomrule
    \end{tabular}
\end{table*}

\subsection{Polarimetric Recovery under Low-Light Noise}
\textbf{Dataset setup.} Low-light conditions typically induce severe photon starvation, resulting in a dramatically reduced SNR. To evaluate our method's robustness against such intensified noise, we utilize the PLIE dataset \cite{zhou2023polarization}, a real-world captured pairwise benchmark comprising $100$ training scenes and $30$ testing scenes. The use of real-captured data ensures that the evaluated noise distributions reflect authentic sensor characteristics rather than simplified synthetic Gaussian approximations.

\textbf{Quantitative comparison.} We compare our unified architecture against state-of-the-art polarimetric low-light enhancement methods, including IPLNet \cite{hu2020iplnet} (OL'20), ColorPolarNet \cite{xu2022colorpolarnet} (TIM'22), and PLIE \cite{zhou2023polarization} (AAAI'23). Quantitative results are presented in \Tref{tab: LowLight}. Our method consistently establishes new state-of-the-art performance across all metrics. These gains demonstrate that unlike existing methods that excel only in specific metrics, our joint image-Stokes processing paradigm achieves comprehensive superiority in overcoming severe observation noise.

\textbf{Qualitative evaluation.} Visual comparisons are provided in \Fref{fig: LowLight}\footnote{Additional qualitative results can be found in the supplementary material.}. Our method exhibits superior robustness in suppressing severe real-world noise while meticulously preserving fine-grained polarimetric structures. For instance, in the enlarged regions of the AoP ($\bm{\theta}$) maps, competing methods, which either rely on single-stage single-domain designs (\eg, IPLNet \cite{hu2020iplnet} and PLIE \cite{zhou2023polarization}) or multi-stage multi-domain architectures (\eg, ColorPolarNet \cite{xu2022colorpolarnet}), fail to exploit synergistic cross-domain cues or suffer from error accumulation. Consequently, they produce chaotic and noisy angular reconstructions. In contrast, our single-stage multi-domain framework leverages the structural priors from the Stokes domain via the CDFM modules, effectively reconstructing clean textural details. Furthermore, regarding the TI ($\mathbf{I}$) images, our approach successfully eliminates color bias, yielding results that are visually most consistent with the ground-truth references.

\begin{figure*}[t]
    \centering
    \includegraphics[width=1.0\linewidth]{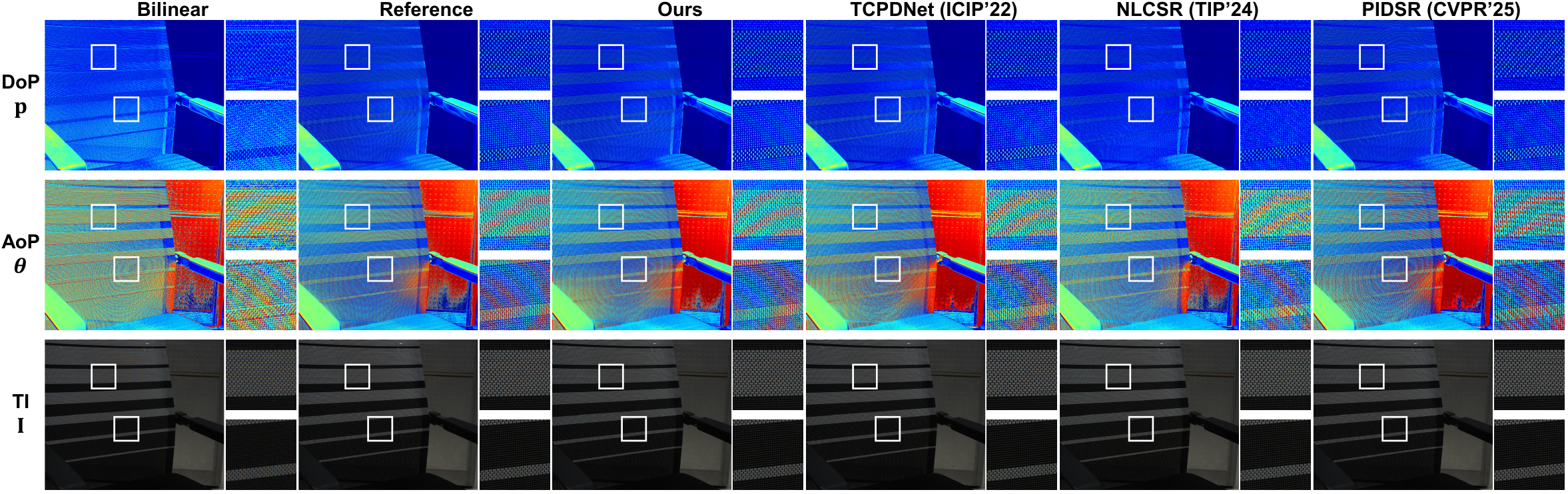}
    \caption{Qualitative comparisons on the PIDSR dataset \cite{zhou2025pidsr} for handling mosaicing artifacts. While existing methods struggle with hallucinated textures or underestimated values in polarimetric parameters, our method recovers physically faithful and artifact-free DoP and AoP structures. Please zoom in for better details.}
    \label{fig: MosaicingArtifacts}
\end{figure*}

\begin{table*}[t]
    \caption{Quantitative comparisons on the PIDSR dataset \cite{zhou2025pidsr} for handling mosaicing artifacts. The best and second-best results are highlighted in \textbf{bold} and \underline{underline}, respectively.}
    \label{tab: MosaicingArtifacts}
    \centering
    \begin{tabular}{lcccccc}
        \toprule
        Method 
        & PSNR-DoP & SSIM-DoP 
        & PSNR-AoP & SSIM-AoP 
        & PSNR-TI  & SSIM-TI \\
        \midrule
        Ours 
        & \textbf{41.09} & \textbf{0.944} 
        & \textbf{16.77} & \textbf{0.510} 
        & \textbf{48.98} & \textbf{0.994} \\
        TCPDNet \cite{nguyen2022two} 
        & 39.15 & 0.929 
        & 15.16 & 0.457 
        & 43.06 & 0.984 \\
        NLCSR \cite{luo2024learning} 
        & 35.75 & 0.891 
        & \underline{16.68} & 0.384 
        & 39.83 & 0.975 \\
        PIDSR \cite{zhou2025pidsr} 
        & \underline{40.42} & \underline{0.939} 
        & 15.15 & \underline{0.465} 
        & \underline{47.53} & \underline{0.992} \\
        Bilinear interpolation
        & 31.08 & 0.813 
        & 14.78 & 0.308 
        & 37.57 & 0.956 \\
        \bottomrule
    \end{tabular}
\end{table*}

\subsection{Polarimetric Recovery under Motion Blur}
\textbf{Dataset setup.} Motion blur is an almost inevitable degradation in dynamic polarimetric imaging scenarios. The inherent light attenuation caused by polarization filters necessitates longer exposure times, which severely exacerbates camera shake or object motion. To evaluate our architecture's ability to resolve such complex spatial degradation, we utilize the PolDeblur dataset \cite{zhou2025learning}, a synthetic pair-wise benchmark comprising $400$ training scenes and $30$ testing scenes. 

\textbf{Quantitative comparison.} Currently, PolDeblur \cite{zhou2025learning} (IJCV'25) stands as the only specialized state-of-the-art method dedicated to deblurring polarized images. To provide a comprehensive analysis, we additionally evaluate two leading polarimetric low-light enhancement models, ColorPolarNet \cite{xu2022colorpolarnet} (TIM'22) and PLIE \cite{zhou2023polarization} (AAAI'23), by retraining them on the motion blur dataset. As reported in \Tref{tab: MotionBlur}, our architecture consistently outperforms all compared approaches across every metric. Meanwhile, ColorPolarNet \cite{xu2022colorpolarnet} and PLIE \cite{zhou2023polarization} exhibit significant performance drops. This drastic contrast empirically proves that existing polarimetric architectures are heavily over-specialized for single degradation types and cannot effectively adapt to spatial blur, whereas our structurally unified backbone maintains robust superiority.

\textbf{Qualitative evaluation.} Visual comparisons on the testing set are provided in \Fref{fig: MotionBlurS}\footnote{Please see the supplementary material for additional results.}. Regarding the TI ($\mathbf{I}$), our method successfully reconstructs high-fidelity textual details. For instance, in the second enlarged region, the text below the character ``D'' is distinctly recovered. In contrast, while the specialized PolDeblur \cite{zhou2025learning} yields sharper edges than the low-light baselines (ColorPolarNet \cite{xu2022colorpolarnet} and PLIE \cite{zhou2023polarization}), it suffers from noticeable ringing artifacts (spurious oscillations near edges) due to its flawed error-accumulating multi-stage design. For the DoP $\mathbf{p}$ and AoP $\bm{\theta}$, our results exhibit minimal structural distortion, accurately restoring the intrinsic polarization states. 

\textbf{Real-captured examples.} More importantly, to validate the practical robustness of our method, we extend our evaluation to real-captured blurred polarized images, as shown in \Fref{fig: MotionBlurR}\footnote{Additional examples are included in the supplementary material.}. Despite being trained exclusively on the synthetic dataset, our framework demonstrates exceptional sim-to-real generalization. It effectively mitigates complex, real-world motion blur and suppresses artifacts in the DoP and AoP, further highlighting the efficacy of our physically constrained joint image-Stokes processing paradigm in real-world deployments.

\begin{table*}[t]
    \caption{Quantitative ablation study of different architectural components and physical constraints on the PolDeblur dataset \cite{zhou2025learning}. The best and second-best results are highlighted in \textbf{bold} and \underline{underline}, respectively.}
    \label{tab: AblationStudy}
    \centering
    \begin{tabular}{llcccccc}
        \toprule
        Types & Model variants & PSNR-DoP & SSIM-DoP & PSNR-AoP & SSIM-AoP & PSNR-TI & SSIM-TI \\
        \hline
        \multirow{2}{*}{Macro-architecture} 
        & W/o Stokes domain & 30.19 & 0.799 & 17.24 & 0.366 & 32.19 & 0.903 \\
        & W/o image domain & 30.28 & \underline{0.801} & 17.91 & 0.367 & 32.53 & 0.906 \\
        \hline
        \multirow{3}{*}{Micro-level modules} 
        & W/o CAFA (replaced with MDTA \cite{zamir2022restormer}) & 30.19 & 0.798 & 17.04 & 0.354 & 32.85 & 0.913 \\
        & W/o CDFM (replaced with GDFN \cite{zamir2022restormer}) & 29.90 & 0.797 & 17.21 & 0.360 & 32.34 & 0.906 \\
        & W/o refinement module & 30.26 & 0.798 & 17.52 & 0.370 & 32.58 & 0.909 \\
        \hline
        \multirow{2}{*}{Objective functions} 
        & W/o physical loss $\mathcal{R}_{\text{i}}$ & \underline{30.30} & \underline{0.801} & \textbf{18.33} & \textbf{0.381} & \underline{33.96} & \underline{0.916} \\
        & W/o physical loss $\mathcal{R}_{\text{s}}$ & 30.28 & 0.799 & \underline{18.06} & \underline{0.379} & 33.18 & 0.910 \\
        \midrule
        \multicolumn{2}{l}{\textbf{Complete method}} & \textbf{30.31} & \textbf{0.804} & 17.92 & 0.378 & \textbf{34.56} & \textbf{0.932} \\
        \bottomrule
    \end{tabular}
\end{table*}

\subsection{Polarimetric Recovery under Mosaicing Artifacts}
\textbf{Dataset setup.} DoFP polarization cameras inherently trade spatial resolution for angular resolution, necessitating a preliminary demosaicing process that frequently introduces severe spatial aliasing and interpolation artifacts. To evaluate our method's efficacy in resolving these artifacts, we utilize the synthetic pairwise PIDSR dataset \cite{zhou2025pidsr}, comprising $108$ training scenes and $30$ testing scenes. For a more rigorous evaluation, we adapt the original demosaicing protocol: instead of using the downsampled $512 \times 612$ images as the demosaicing reference, we directly employ the full-resolution $2048 \times 2448$ images to ensure assessment on complete-resolution polarimetric states. Furthermore, to maximize practical applicability, our network does not require complex, task-specific raw-data ingestion modules. Instead, we directly use bilinear interpolation, \ie, the default demosaicing scheme in commercial polarization cameras, to generate the initial inputs, relying entirely on our unified backbone to rectify the subsequent artifacts.

\textbf{Quantitative comparison.} We benchmark our unified architecture against state-of-the-art specialized polarimetric demosaicing methods, including TCPDNet \cite{nguyen2022two} (ICIP'22), NLCSR \cite{luo2024learning} (TIP'24), and PIDSR \cite{zhou2025pidsr} (CVPR'25), along with bilinear interpolation. As detailed in \Tref{tab: MosaicingArtifacts}, our method consistently outperforms all compared approaches across every metric. These results powerfully demonstrate that our unified architecture, operating on simple bilinearly interpolated inputs, can surpass highly specialized models engineered specifically for demosaicing tasks.

\textbf{Qualitative evaluation.} Visual comparisons are provided in \fref{fig: MosaicingArtifacts}\footnote{More qualitative results can be found in the supplementary material.}. While all evaluated methods achieve comparable, satisfactory performance on the TI ($\mathbf{I}$), the recovery of polarimetric parameters exposes significant vulnerabilities in existing approaches. Our approach effectively suppresses mosaicing artifacts, yielding continuous and physically consistent structures. For example, in the second enlarged region of the DoP ($\mathbf{p}$) maps, specialized deep learning-based methods (TCPDNet \cite{nguyen2022two} and PIDSR \cite{zhou2025pidsr}) hallucinate spurious high-frequency textures in inherently low-frequency areas. Conversely, the dictionary-learning approach (NLCSR \cite{luo2024learning}) tends to severely underestimate the polarization magnitudes. In contrast, our results are visually and structurally much closer to the ground-truth reference, heavily benefiting from the joint image-Stokes modeling paradigm that explicitly avoids multi-stage error accumulation.

\subsection{Ablation Studies}
To validate the effectiveness of our core architectural designs and physical constraints, we conduct a comprehensive series of ablation studies. Given that our framework strictly shares the exact same topology across all degradation tasks, we select the highly challenging motion blur scenario (\ie, the PolDeblur dataset \cite{zhou2025learning}) as a representative benchmark to avoid redundant enumerations. As detailed in \Tref{tab: AblationStudy}, we incrementally isolate and analyze the contributions of our proposed method from three distinct dimensions: the macro-architecture, the micro-level modules, and the objective functions.

\textbf{Macro-architecture (dual-domain necessity).} We first examine our single-stage multi-domain framework by comparing it  against two degenerated single-domain variants: one operating exclusively in the image domain (w/o Stokes domain) and another relying solely on the Stokes parameters (w/o image domain). Both variants exhibit severe performance degradation, particularly in the TI ($\mathbf{I}$) metric. This validates our fundamental hypothesis: neither domain alone provides sufficient representational capacity, and exploiting the complementary synergistic cues between them is indispensable for high-fidelity restoration.

\textbf{Micro-level modules (cross-domain interaction).} Next, we highlight the structural significance of the proposed CDCI units. We replace our CAFA and CDFM components with the MDTA (Multi-Dconv Head Transposed Attention) and GDFN (Gated-Dconv Feed-forward Network) blocks introduced by Restormer \cite{zamir2022restormer}, which originally inspired our module layout. As shown in \Tref{tab: AblationStudy}, replacing CAFA causes a notable drop in AoP ($\bm{\theta}$), while removing CDFM severely compromises DoP ($\mathbf{p}$). This occurs because the blocks introduced by Restormer \cite{zamir2022restormer}, although highly effective for single-domain image restoration, are not equipped to facilitate explicit cross-domain feature interaction and physical modulation. Furthermore, removing the terminal refinement module (w/o refinement) also leads to inferior performance across all metrics, confirming its necessity for fine-grained texture recovery.

\textbf{Objective functions (physical constraints).} Finally, we analyze the impact of the proposed physical consistency constraints ($\mathcal{R}_{\text{i}}$ and $\mathcal{R}_{\text{s}}$). As evident in \Tref{tab: AblationStudy}, removing either $\mathcal{R}_{\text{i}}$ or $\mathcal{R}_{\text{s}}$ introduces mild metric-level trade-offs. For instance, without the image-domain physical constraint $\mathcal{R}_{\text{i}}$, the model achieves a slightly higher AoP PSNR but suffers a significant performance degradation in TI. Ultimately, the full model yields the most balanced and comprehensive performance across all three parameters (DoP, AoP, and TI). This observation indicates that the dual-branch cross-domain architectural design remains the primary contributor, while the physical consistency losses serve as essential regularizers that explicitly prevent the network from over-optimizing a single parameter at the expense of overall physical equilibrium.

\begin{figure*}[t]
    \centering
    \includegraphics[width=1.0\linewidth]{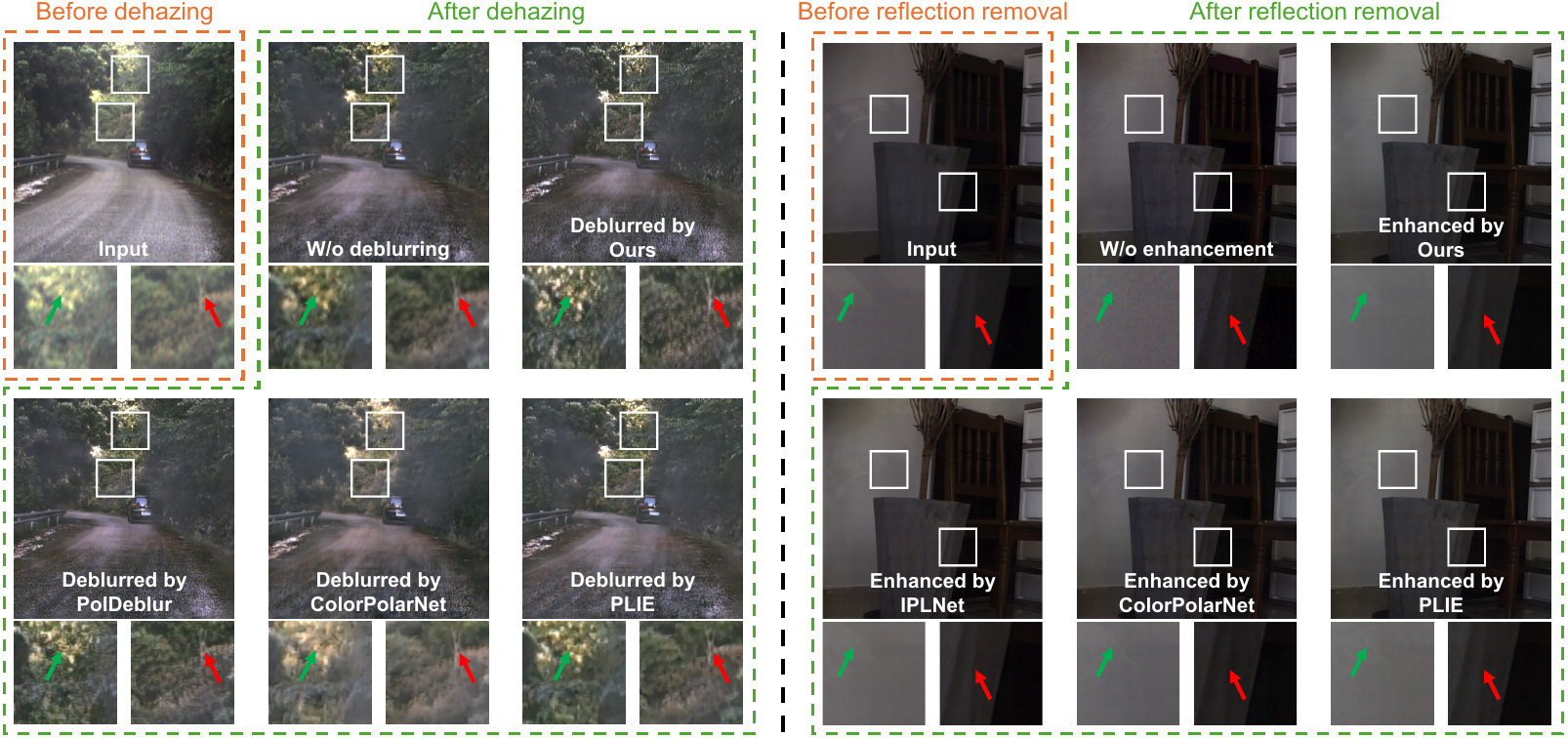}
    \caption{Impact of polarimetric restoration on downstream physics-based vision applications. For visual simplicity, only the TI ($\mathbf{I}$) is presented. Left: Polarization-guided dehazing \cite{zhou2021learning} under motion blur in a driving scenario. Right: Polarization-guided reflection removal \cite{lyu2022physics} under low-light noise in an indoor scenario (the input is digitally amplified for visualization). Please zoom in for better details.}
    \label{fig: Application}
\end{figure*}

\subsection{Application}
To demonstrate the practical significance of our unified architecture, we evaluate how the high-quality polarimetric representations recovered by our method can facilitate and improve downstream physics-based vision applications. Since downstream algorithms heavily rely on the accuracy of physical parameters (\ie, DoP and AoP) to formulate physical constraints, severe degradations can cause these models to fail.

\textbf{Deblurring for polarization-guided dehazing.} As an initial example, we evaluate the impact of deblurring on image dehazing \cite{zhou2021learning} in dynamic driving scenarios. As shown in \Fref{fig: Application} (left), the input observation is heavily compromised by both atmospheric haze and motion blur (for simplicity, only the TI $\mathbf{I}$ is visualized). Directly applying the polarization-based dehazing method \cite{zhou2021learning} to the degraded input fails to estimate the correct scattering model, leaving prominent haze residuals. When prior restoration is applied, competing methods exhibit significant drawbacks: the specialized PolDeblur \cite{zhou2025learning} hallucinates non-existent structures (indicated by the circle-like artifact, green arrow), while the low-light-oriented ColorPolarNet \cite{xu2022colorpolarnet} and PLIE \cite{zhou2023polarization} over-smooth the high-frequency textural details (e.g., the tree branch, red arrow). In contrast, our method faithfully recovers the sharp underlying polarimetric cues, enabling the downstream dehazing algorithm to yield a clean, artifact-free, and visually pleasing result.

\textbf{Low-light enhancement for reflection removal.} Furthermore, we investigate the impact of low-light enhancement on polarization-guided reflection removal \cite{lyu2022physics} in indoor scenarios. As depicted in \Fref{fig: Application} (right), the input is corrupted by both glass reflections and severe low-light noise (note that the input has been digitally amplified for visualization purposes, as the original captures are nearly entirely dark). Directly performing reflection removal \cite{lyu2022physics} without enhancement drastically amplifies the background noise, rendering the transmission layer unrecognizable. When applying existing low-light enhancement methods as a preprocessing step, IPLNet \cite{hu2020iplnet} and ColorPolarNet \cite{xu2022colorpolarnet} fail to preserve the crucial polarization differences between reflected and transmitted light, resulting in severe residual reflection artifacts on the wall (green arrow). Meanwhile, PLIE \cite{zhou2023polarization} leaves noticeable noise-like textural patterns on the flowerpot (red arrow). Our method preserves the physical polarization states even under extreme noise, thereby allowing the downstream algorithm to cleanly separate the transmission layer with minimal artifacts.

\subsection{Discussion on General-Purpose RGB Models}
To justify the necessity of designing a specialized, polarization-aware architecture, we compare our framework against Restormer \cite{zamir2022restormer}, a classical general-purpose RGB restoration model that also conceptually inspired our micro-level attention designs. To ensure a comprehensive analysis, we evaluate two distinct configurations across all three degradation datasets: (1) \textit{Original version}: The standard architecture that processes the four polarized images independently in a frame-by-frame manner, as it was inherently designed for single-image inputs; (2) \textit{Modified version}: We adapt its layers to ingest and reconstruct all polarized images simultaneously. Note that for a fair comparison, we apply the same loss function as ours when retraining.

As reported in \Tref{tab: RGBModel}, the original version yields sub-optimal polarimetric parameters. This exposes the fatal flaw of polarization-unawareness: processing spatial intensities independently destroys the intrinsic physical correlations required to accurately derive Stokes vectors. Conversely, while the modified version benefits significantly from simultaneous processing and dual-domain loss supervision, it still falls consistently short of our framework across all metrics and datasets. This performance gap empirically validates a crucial insight: simply appending dual-domain loss functions to a general-purpose vision backbone is insufficient. Explicitly modeling cross-domain physical interactions via our designed joint image-Stokes representation and CDCI modules is indispensable for high-fidelity polarimetric recovery.

\begin{table}[t]
    \caption{Quantitative comparison with general-purpose RGB restoration models. Our unified framework consistently outperforms both the original and modified versions of Restormer \cite{zamir2022restormer}.}
    \label{tab: RGBModel}
    \centering
    \resizebox{1.0\linewidth}{!}{
    \begin{tabular}{lccc}
        \toprule
        Method & DoP (PSNR / SSIM) & AoP (PSNR / SSIM) & TI (PSNR / SSIM) \\
        \midrule
        \multicolumn{4}{c}{Low-Light Noise (PLIE dataset \cite{zhou2023polarization})} \\
        \midrule
        Restormer \cite{zamir2022restormer} (Original) & 26.42 / 0.761 & 15.91 / 0.328 & 38.50 / 0.975 \\
        Restormer \cite{zamir2022restormer} (Modified) & 28.97 / 0.816 & 16.35 / 0.332 & 40.12 / 0.984 \\
        Ours                 & 30.61 / 0.842 & 16.69 / 0.446 & 40.78 / 0.986 \\
        \midrule
        \multicolumn{4}{c}{Motion Blur (PolDeblur dataset \cite{zhou2025learning})} \\
        \midrule
        Restormer \cite{zamir2022restormer} (Original) & 28.69 / 0.763 & 16.93 / 0.307 & 28.34 / 0.864 \\
        Restormer \cite{zamir2022restormer} (Modified) & 29.06 / 0.771 & 17.01 / 0.334 & 31.83 / 0.898 \\
        Ours                 & 30.31 / 0.804 & 17.92 / 0.378 & 34.56 / 0.932 \\
        \midrule
        \multicolumn{4}{c}{Mosaicing Artifacts (PIDSR dataset \cite{zhou2025pidsr})} \\
        \midrule
        Restormer \cite{zamir2022restormer} (Original) & 40.08 / 0.930 & 14.99 / 0.442 & 47.10 / 0.982 \\
        Restormer \cite{zamir2022restormer} (Modified) & 40.29 / 0.937 & 15.08 / 0.457 & 47.49 / 0.990 \\
        Ours                 & 41.09 / 0.944 & 16.77 / 0.510 & 48.98 / 0.994 \\
        \bottomrule
    \end{tabular}
    }
\end{table}

\section{Conclusion}
We propose a structurally unified, single-stage multi-domain framework for polarimetric image restoration. By introducing CDCI units as the core building blocks, our method intrinsically fuses image- and Stokes-domain representations without multi-stage error accumulation. Utilizing the exact same network topology, this unified backbone consistently achieves state-of-the-art performance across diverse degradations, including low-light noise, motion blur, and mosaicing artifacts, and effectively enhances downstream physics-based vision applications.

\textbf{Limitations.} Currently, our framework focuses on static image restoration; extending it to polarimetric video requires explicitly modeling temporal consistencies to prevent physical flickering. Furthermore, consistent with mainstream commercial DoFP sensors, our method addresses only linear polarization. Incorporating circular polarization would necessitate customized acquisition systems and expanded physical regularizations.


{\small
\bibliographystyle{ieee_fullname}
\bibliography{egbib}
}

\vspace{-10mm}

\begin{IEEEbiography}
[{\includegraphics[width=1in,height=1.25in,clip,keepaspectratio]{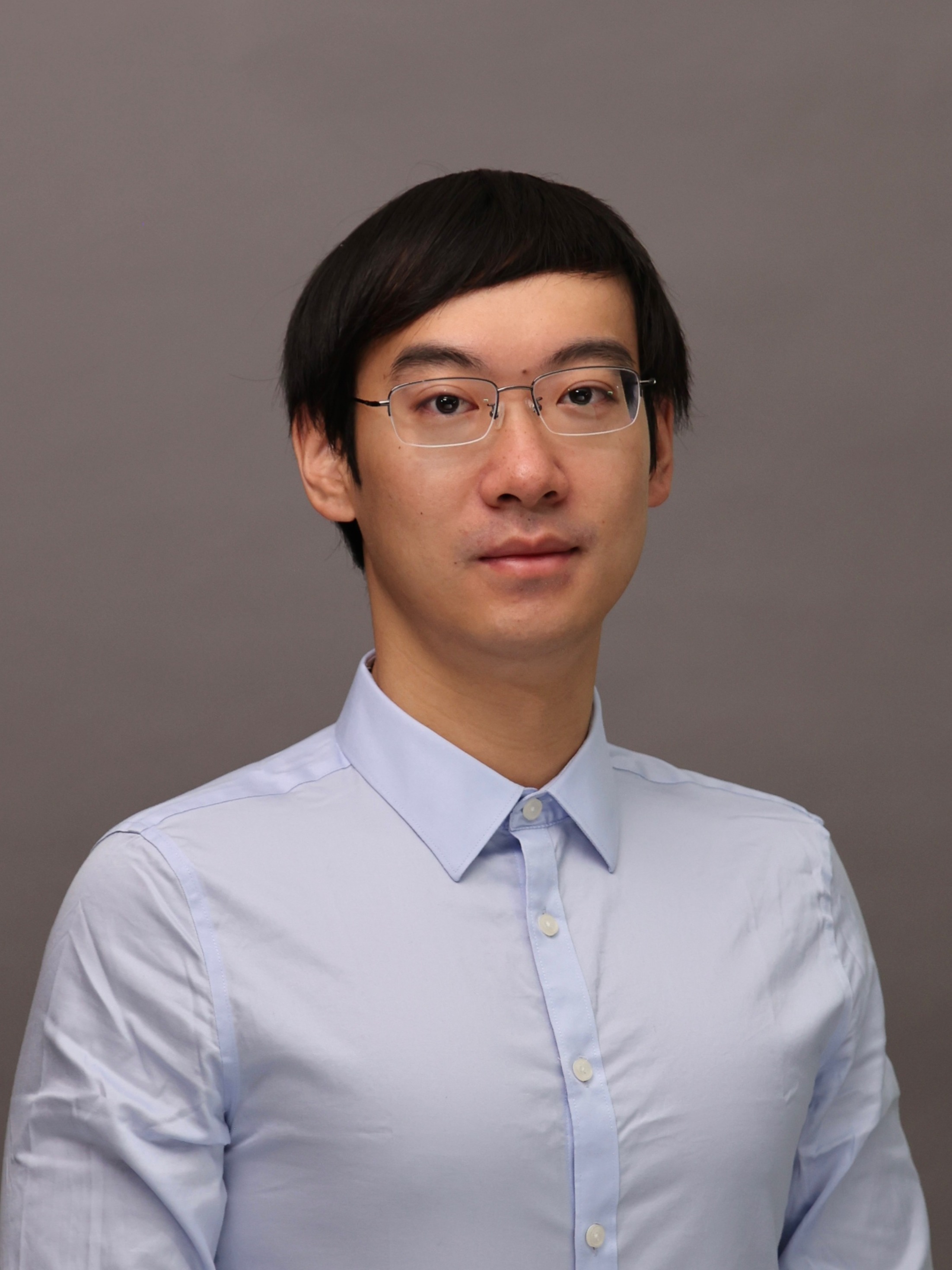}}]
{Chu Zhou} received the B.E. degree from Huazhong University of Science and Technology in 2019 and the Ph.D. degree from School of Intelligence Science and Technology, Peking University in 2024. He is currently an assistant professor by special appointment at Digital Contents and Media Sciences Research Division, National Institute of Informatics. His research interest includes computational photography and computer vision, with a focus on unconventional camera-based vision.
\end{IEEEbiography}
\vspace{-10mm}

\begin{IEEEbiography}
[{\includegraphics[width=1in,height=1.25in,clip,keepaspectratio]{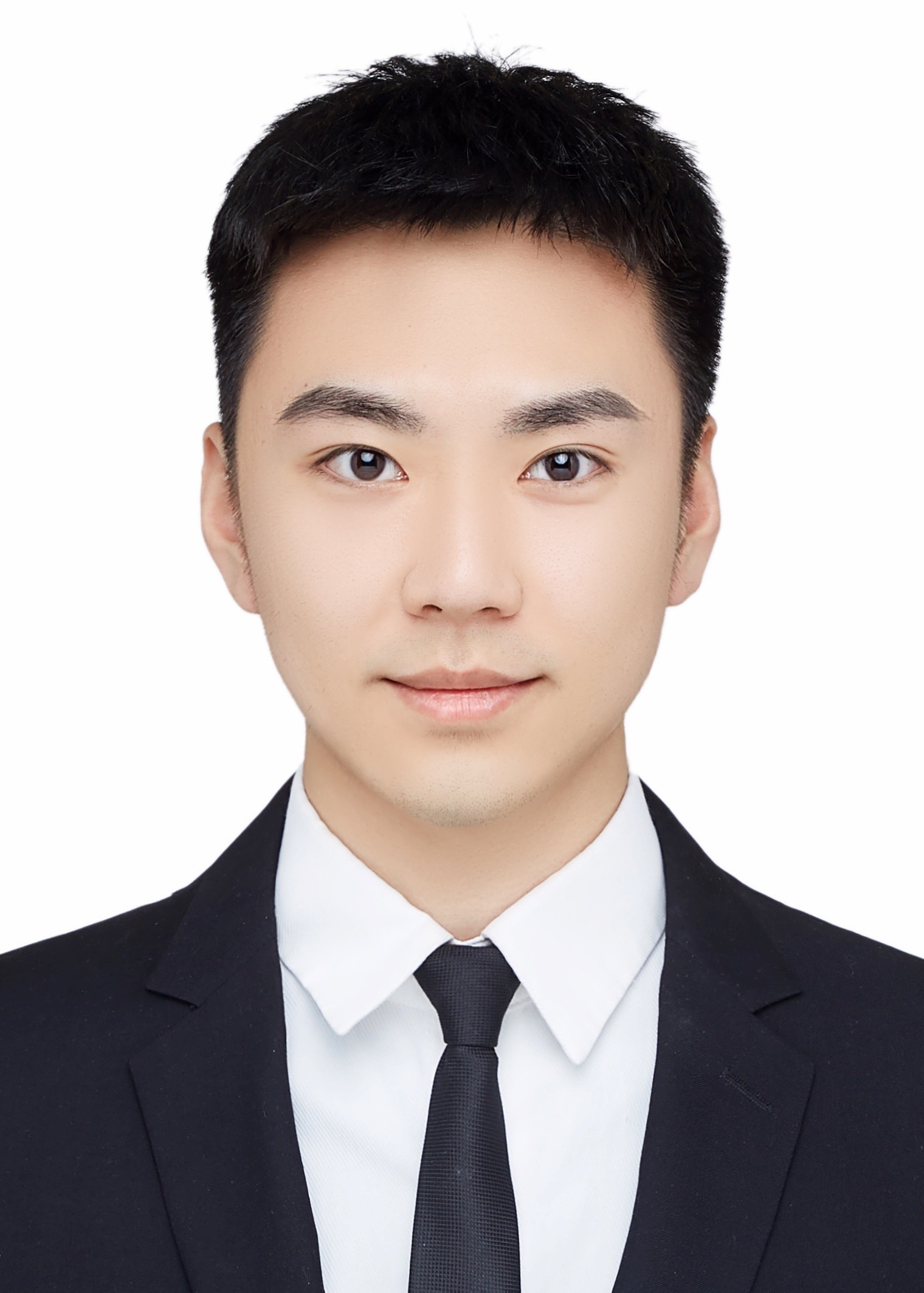}}]
{Yufei Han} received a B.S. degree in Communication Engineering from the Beijing University of Posts and Telecommunications, Beijing, China in 2022. He is currently working toward a Ph.D. degree in Artificial Intelligence with the School of Artificial Intelligence, Beijing University of Posts and Telecommunications, Beijing, China. His research interests include polarization-based vision and 3D reconstruction.
\end{IEEEbiography}
\vspace{-10mm}

\begin{IEEEbiography}
[{\includegraphics[width=1in,height=1.25in,clip,keepaspectratio]{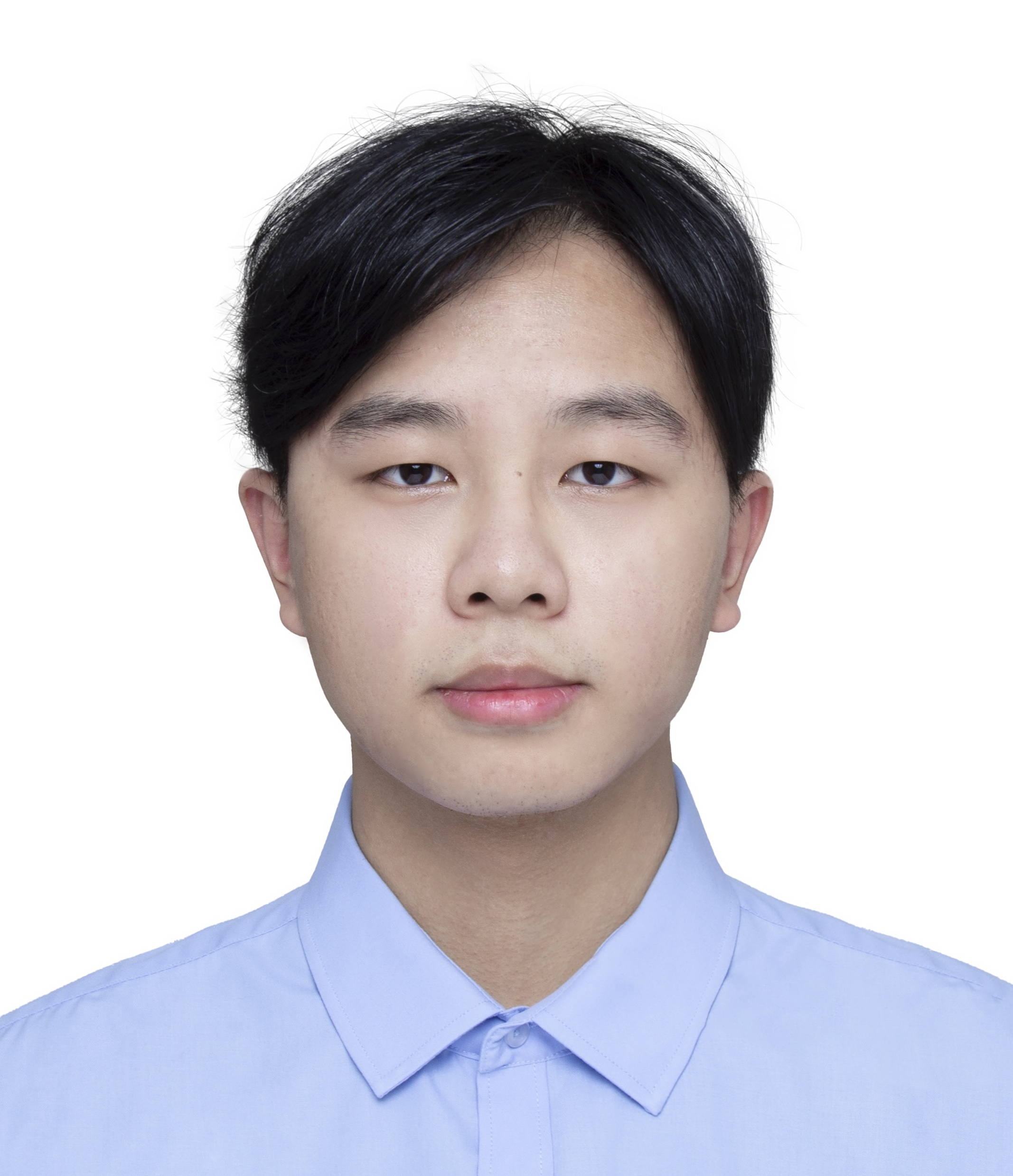}}]
{Junda Liao} received the bachelor’s degree in software engineering from Nanjing University, Nanjing, China, in 2020, and the master’s degree in software engineering from Nanjing University, Nanjing, China, in 2022. He is currently working toward the doctoral degree with the Department of Computer Science, Graduate School of Information Science and Technology, The University of Tokyo, Tokyo, Japan, under the supervision of Prof. Imari Sato. He is also a research intern with the National Institute of Informatics, Tokyo, Japan. His research interests include image processing and computer vision.
\end{IEEEbiography}
\vspace{-10mm}

\begin{IEEEbiography}
[{\includegraphics[width=1in,height=1.25in,clip,keepaspectratio]{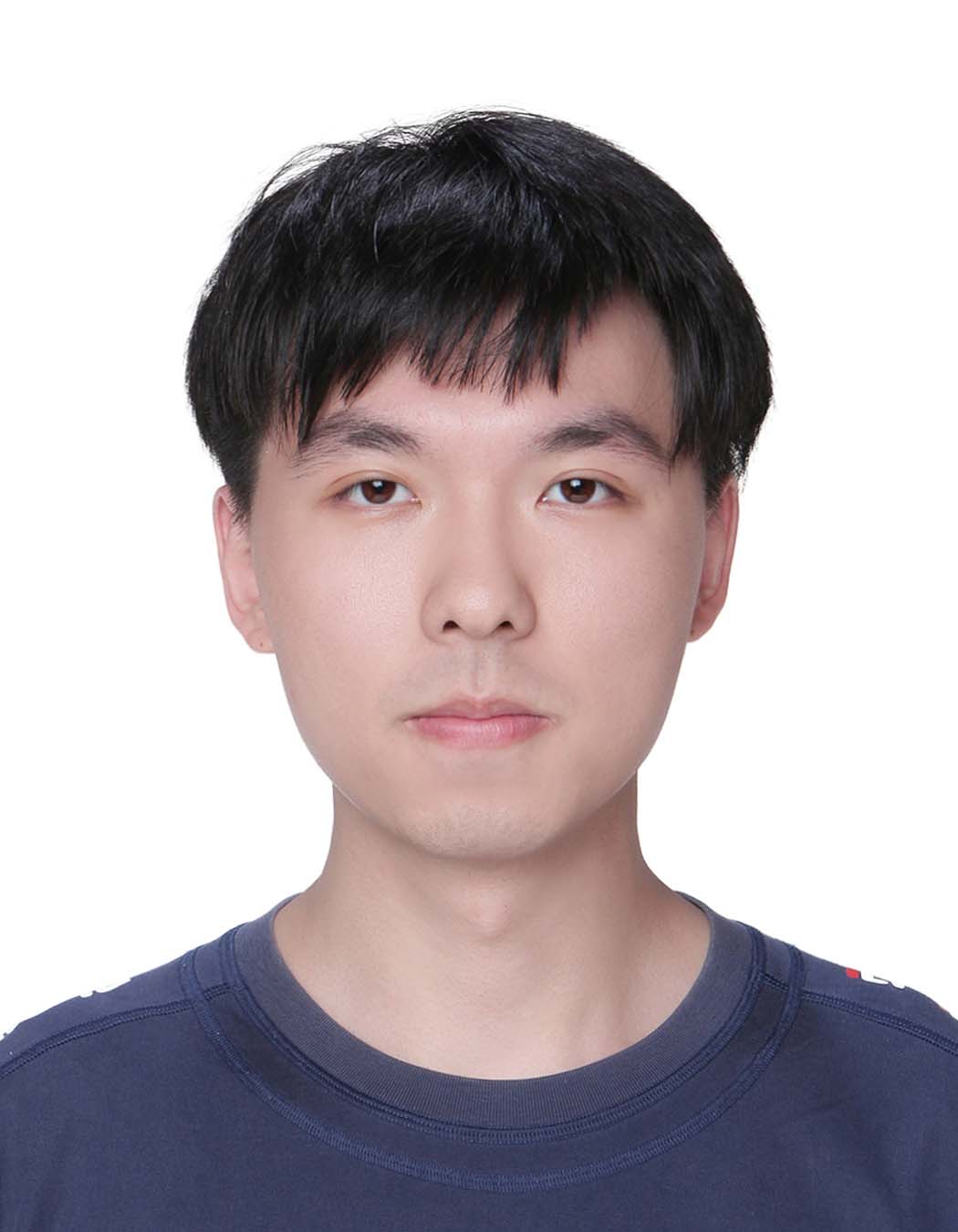}}]
{Linrui Dai} received his M.S. degree in Computer Science from Shanghai Jiao Tong University in 2024. Currently, he is working toward his Ph.D. degree in the Graduate School of Information Science and Technology, The University of Tokyo, Japan. He is also affiliated with the National Institute of Informatics, Japan as a research intern from 2025. His research interests include machine learning, computer vision and computed tomography.
\end{IEEEbiography}
\vspace{-10mm}

\begin{IEEEbiography}
[{\includegraphics[width=1in,height=1.25in,clip,keepaspectratio]{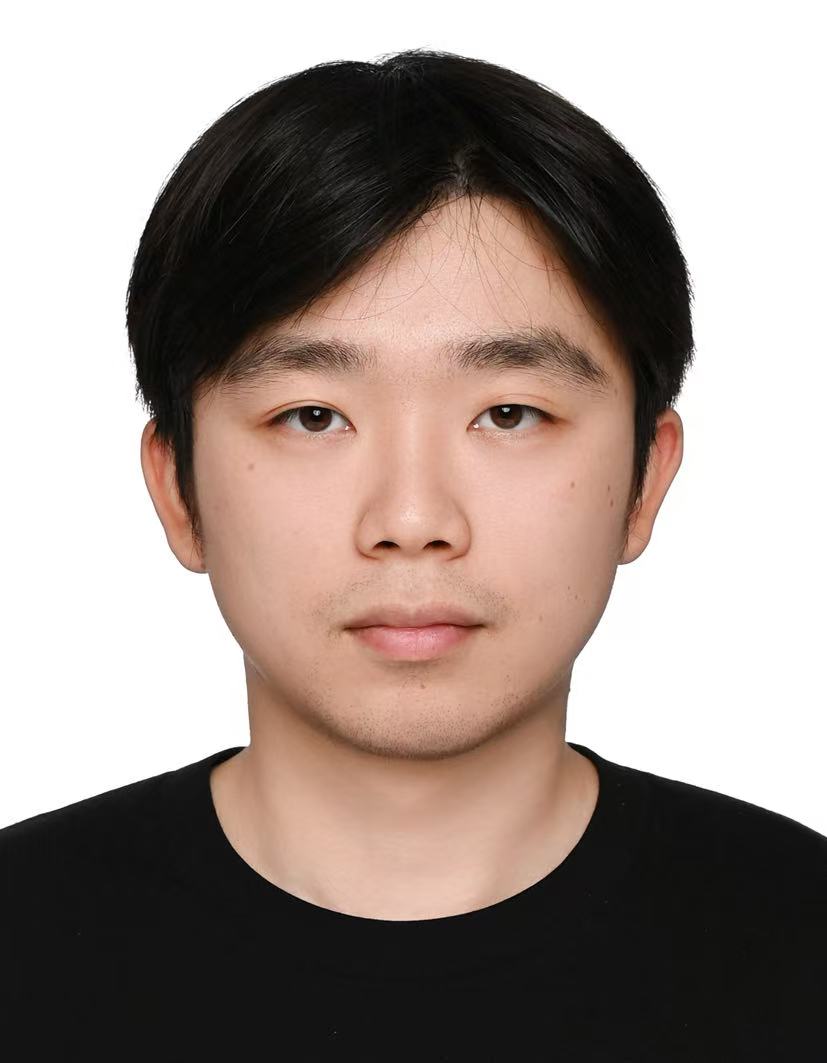}}]
{Wangze Xu} received the B.Eng. degree from Sichuan University in 2022 and the M.Sc. degree from Peking University in 2025. He is currently working toward the Ph.D. degree with the Department of Computer Science, The University of Tokyo, under the supervision of Prof. Imari Sato. He is also a Research Intern with the National Institute of Informatics (NII), Tokyo, Japan. His research interests include 3D reconstruction and neural rendering.
\end{IEEEbiography}
\vspace{-10mm}

\begin{IEEEbiography}
[{\includegraphics[width=1in,height=1.25in,clip,keepaspectratio]{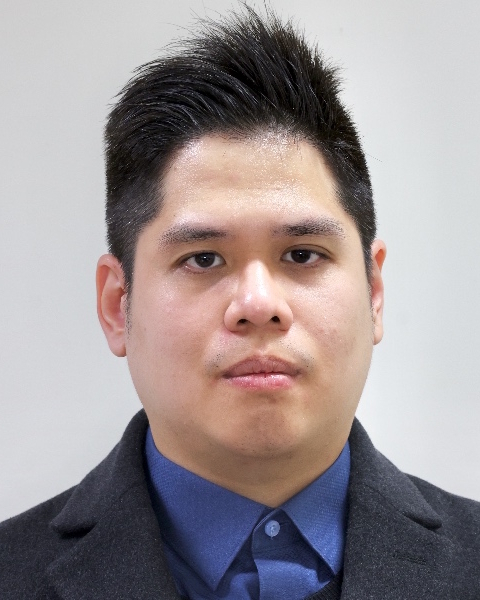}}]
{Art Subpa-Asa} received the BE and ME degrees in computer engineering from Chulalongkorn University, Bangkok, Thailand, in 2009 and 2011, respectively, and the PhD degree in information processing from the Tokyo Institute of Technology, Tokyo, Japan, in 2018. He is currently a researcher with the National Institute of Informatics. His research interests include computational photography of global illumination and acquisition system.
\end{IEEEbiography}
\vspace{-10mm}

\begin{IEEEbiography}
[{\includegraphics[width=1in,height=1.25in,clip,keepaspectratio]{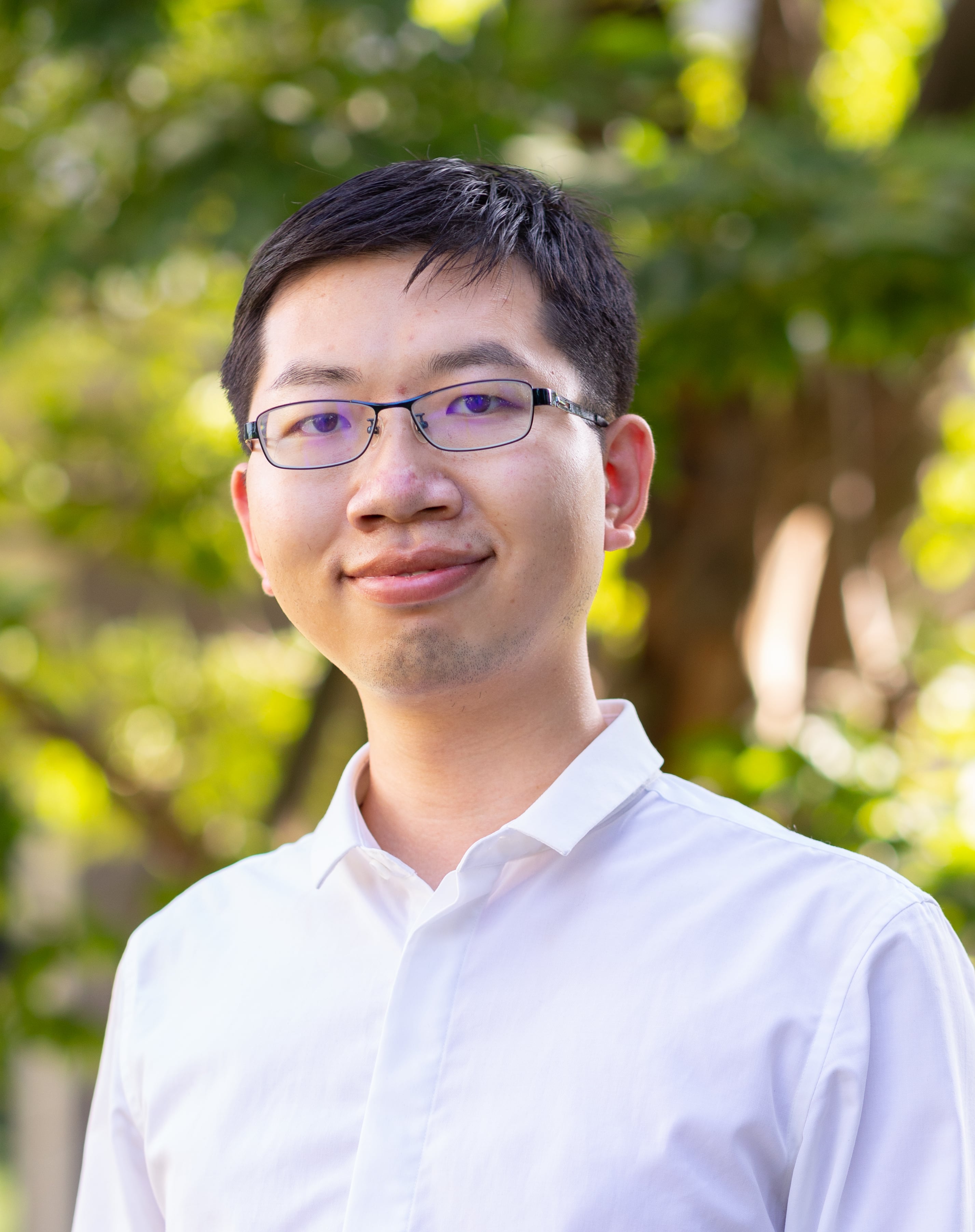}}]
{Heng Guo} received the BE and ME degrees from University of Electronic Science and Technology, and the PhD degree from Osaka University, in 2015, 2018, and 2022. He is currently a specially-appointed research Professor at Beijing University of Posts and Telecommunications (BUPT). Before joining BUPT, he was a specially-appointed assistant professor at Osaka University from 2022 to 2023. His research interest includes computational photography, physics-based computer vision, and 3D reconstruction. He served as reviewers of CVPR, ICCV, ECCV, TPAMI, IJCV.
\end{IEEEbiography}
\vspace{-10mm}

\begin{IEEEbiography}
[{\includegraphics[width=1in,height=1.25in,clip,keepaspectratio]{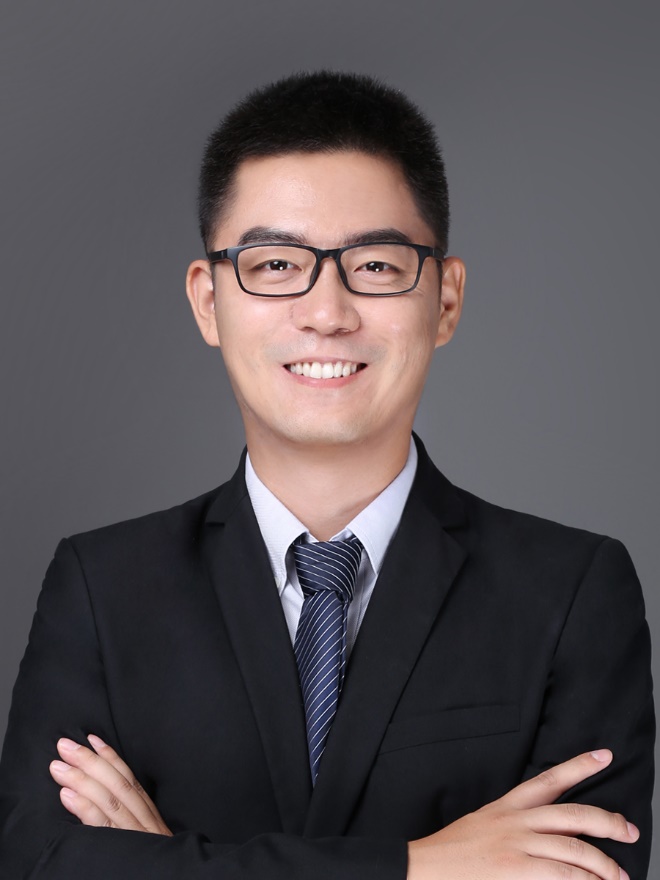}}]
{Boxin Shi} received the B.E. degree from the Beijing University of Posts and Telecommunications, the M.E. degree from Peking University, and the Ph.D. degree from the University of Tokyo, in 2007, 2010, and 2013. He is currently a Boya Young Fellow Associate Professor (with tenure) and Research Professor at Peking University, where he leads the Camera Intelligence Lab. Before joining PKU, he did research with MIT Media Lab, Singapore University of Technology and Design, Nanyang Technological University, National Institute of Advanced Industrial Science and Technology, from 2013 to 2017. His papers were awarded as Best Paper, Runners-Up at CVPR 2024, ICCP 2015, and selected as Best Paper candidate at ICCV 2015. He is an associate editor of TPAMI/IJCV and an area chair of CVPR/ICCV/ECCV. He is a senior member of IEEE.
\end{IEEEbiography}
\vspace{-10mm}

\begin{IEEEbiography}
[{\includegraphics[width=1in,height=1.25in,clip,keepaspectratio]{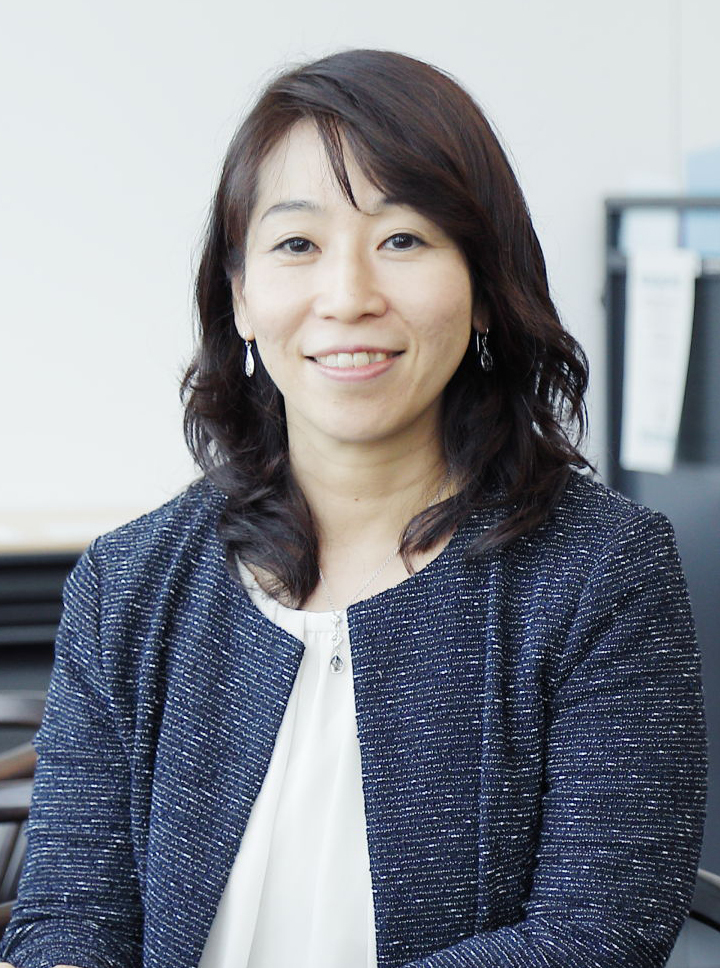}}]
{Imari Sato} received the B.S. degree in policy management from Keio University, Tokyo, Japan, in 1994, and the M.S. and Ph.D. degrees in interdisciplinary Information Studies from the University of Tokyo, Tokyo, in 2002 and 2005, respectively. She was a visiting scholar with the Robotics Institute of Carnegie Mellon University, Pittsburgh, PA, USA. In 2005, she joined the National Institute of Informatics, where she is currently a professor/director of the Digital Contents and Media Sciences Research Division. She is a professor at the University of Tokyo and a visiting professor at the Tokyo Institute of Technology, Tokyo, Japan. Her primary research interests include physics-based vision, spectral analysis, image-based modeling, and medical image analysis. She received various research awards, including the Young Scientists’ Prize from the Commendation for Science and Technology by the Minister of Education, Culture, Sports, Science and Technology in 2009, and the Microsoft Research Japan New Faculty Award in 2011.
\end{IEEEbiography}

\end{document}